\documentclass[12pt,a4paper,twoside]{article}

\usepackage{fancyhdr}
\usepackage{amsmath,amsfonts}

\DeclareMathOperator{\trace}{Tr}
\DeclareMathOperator{\sign}{sign}
\DeclareMathOperator{\rpart}{Re}

\newcommand\rmd{\, \mathrm{d}}
\newcommand\rme{\mathrm{e}}
\newcommand\rmi{\mathrm{i}}
\newcommand\G{\mathrm{G}}
\newcommand\bA{\overline{A}}
\newcommand\bB{\overline{B}}
\newcommand{\abs}[1]{\lvert #1 \rvert}
\newcommand{\ket}[1]{\left | #1 \right \rangle}
\newcommand{\bra}[1]{\left \langle #1 \right |}

\newcommand{\GM}{\mathrm{G}_{(\sigma_1,\sigma_2)}(\tilde{N})}
\newcommand{\nZ}{Z_N^{(\sigma_1,\sigma_2)}}
\newcommand{\UN}{\mathrm{U}(N)}
\newcommand{\ONen}{\mathrm{O}^{-}(2N +2)}
\newcommand{\ONep}{\mathrm{O}^{+}(2N)}
\newcommand{\ONo}[1]{\mathrm{O}^{#1}(2N + 1)}
\newcommand{\SpN}{\mathrm{Sp}(2N)}
\newcommand{\SptN}{\mathrm{Sp}(2N)}
\newcommand{\EgU}{E^{\mathrm{CUE}}_N}
\newcommand{\EgC}{E^{{\rm CyUE}}_N}
\newcommand{\vphi}{\boldsymbol{\phi}}
\newcommand{\vpsi}{\boldsymbol{\psi}}
\newcommand{\gsk}{\ket{\boldsymbol{\Psi}_{{\rm g}}}}
\newcommand{\gsb}{\bra{\boldsymbol{\Psi}_{{\rm g}}}}
\newcommand{\lamp}{(\sigma_1,\sigma_2)}

\numberwithin{equation}{section}

\begin{document}
\title{Random Matrix Theory and Entanglement in Quantum Spin Chains}
\author{J.P.~Keating\footnote{e-mail:
j.p.keating@bristol.ac.uk}~
 and F.~Mezzadri\footnote{e-mail: f.mezzadri@bristol.ac.uk} \\
School of Mathematics\\ University of Bristol\\ Bristol BS8 1TW\\
UK}

\date{5 August 2004}

\maketitle
\begin{abstract}
We compute the entropy of entanglement in the ground states of a
general class of quantum spin-chain Hamiltonians --- those that
are related to quadratic forms of Fermi operators --- between the
first $N$ spins and the rest of the system in the limit of
infinite total chain length. We show that the entropy can be
expressed in terms of averages over the classical compact groups
and establish an explicit correspondence between the symmetries of
a given Hamiltonian and those characterizing the Haar measure of
the associated group. These averages are either Toeplitz
determinants or determinants of combinations of Toeplitz and
Hankel matrices. Recent generalizations of the Fisher-Hartwig
conjecture are used to compute the leading order asymptotics of
the entropy as $N\rightarrow\infty$. This is shown to grow
logarithmically with $N$. The constant of proportionality is
determined explicitly, as is the next (constant) term in the
asymptotic expansion.  The logarithmic growth of the entropy was
previously predicted on the basis of numerical computations and
conformal-field-theoretic calculations. In these calculations the
constant of proportionality was determined in terms of the central
charge of the Virasoro algebra.  Our results therefore lead to an
explicit formula for this charge.  We also show that the entropy
is related to solutions of ordinary differential equations of
Painlev\'e type. In some cases these solutions can be evaluated to
all orders using recurrence relations.
\end{abstract}

\vspace{.25cm}

\hspace{.26cm} Mathematics Subject Classification: 81P68, 15A52,
82B20

\newpage

\section{Introduction}
\label{intro}

Random Matrix Theory, as developed by, amongst others, Freeman
Dyson, provides the natural framework for calculating statistical
properties of quantum fluctuations.  It has had a profound impact
in many of the traditional application areas of quantum mechanics,
including condensed matter physics and nuclear physics. Our
purpose here is to establish a new application: to calculating
entanglement in quantum spin chains.

The importance of entangled quantum states lies in their ability
to exhibit correlations that cannot be accounted for classically.
This feature of quantum mechanics has been known to physicists for
almost seven decades --- since Einstein, Podolsky and Rosen
introduced their famous gedankenexperiment. However, it is only
relatively recently that entanglement has come to be viewed as a
physical resource for manipulating quantum information. As for any
other physical resource, like energy or entropy, it important to
be able to quantify entanglement; that is, to assign a measure to
it. This aspect of entangled states is still poorly understood,
especially when the entanglement is shared between more than two
systems. However, when the entanglement of a pure state is shared
between two parties, i.e.~in a {\it bipartite system}, Bennett
{\em et al}~\cite{BBPS96} have shown that it is consistent to
define it as the von Neumann entropy of either of the two parts.

We consider here the general class of quantum spin chains arising
from quadratic chains of fermionic operators in their ground
state.  These systems are partitioned into two contiguous
subchains. If the ground state is non-degenerate, this subdivision
creates a pure bipartite system and we investigate its
entanglement of formation.

As is well known, the systems we are studying exhibit quantum
phase transitions.  These manifest themselves as qualitative
changes in the decay of correlations: algebraic in the proximity
of a critical point and exponential decay away from it.
Entanglement plays a fundamental role in the quantum phase
transitions that occur in interacting lattice systems at zero
temperature~\cite{AOPFP04,VLRK02,JK03,Kor04,CC04}. Under these
conditions the system is in the ground state, which is also a pure
state, and any correlations must be a consequence of the fact the
ground state is entangled. It follows immediately that the
entanglement changes qualitatively in the proximity of critical
points.

Vidal {\it et al}~\cite{VLRK02} studied the ground states of the
XY and XXZ models, partitioning them into two consecutive
subchains; they observed numerically that, when the Hamiltonian
undergoes a phase transition, the entanglement of formation of
this bipartite system grows logarithmically with the size $N$ of
one of the two parts. Jin and Korepin~\cite{JK03} considered the
XX model, which corresponds to an XY spin chain with an isotropic
interaction, and expressed the von Neumann entropy in terms of a
Toeplitz determinant, which they evaluated asymptotically using
the Fisher-Hartwig conjecture~\cite{FH68}.  Their result coincides
with the numerical observations of Vidal {\it et
al}~\cite{VLRK02}, in that they found that the entropy grows like
$\frac{1}{3}\log_2 N$. Subsequently Korepin~\cite{Kor04} and
Calabrese and Cardy~\cite{CC04} showed, using conformal
field-theoretic arguments, that the logarithmic divergence of the
entanglement in one dimensional systems is a general consequence
of the logarithmic growth of the entropy with the size of the
system at phase transitions. These arguments determine the
constant multiplying the leading order $\log_2 N$ term in the
asymptotics to be one-third of the central charge of the
associated Virasoro algebra.

Our goal here is to show that if a quantum spin-chain Hamiltonian
posses certain symmetries, the entanglement can be expressed as an
average over one of the classical compact groups equipped with
Haar measure, i.e. one of the following groups: $\mathrm{U}(N)$,
$\mathrm{Sp}(2N)$ and $\mathrm{O}^\pm(N)$, where the superscript
$\pm$ indicates the connected component of the orthogonal group
with determinant $\pm 1$. Indeed, we establish a precise
correspondence between the functional form of the appropriate Haar
measure and the symmetries of the Hamiltonian. The XX model turns
out to be an example of a system with $\mathrm{U}(N)$ symmetry.
The averages that occur can be expressed either as Toeplitz
determinants, in the case of $\UN$, or as determinants of specific
combinations of Toeplitz and Hankel matrices for the other compact
groups. Recently, Basor and Ehrhardt~\cite{BE02} and Forrester and
Frankel~\cite{FF04} have computed asymptotic formulae for such
determinants, and these allow us to write down the leading-order
and next-to-leading-order terms in the asymptotics of the
entanglement in the limit as the total number of spins tends to
infinity and then as $N\rightarrow\infty$.

We find that in the proximity of a critical point the entanglement
grows logarithmically with $N$, in agreement with the prediction
of Korepin~\cite{Kor04} and Calabrese and Cardy~\cite{CC04}. We
derive a general formula for the associated constant of
proportionality. This is a rational number, the numerator of which
is shown to factorize into a universal part, related to symmetries
of the quantum Hamiltonian and which can be calculated from the
random-matrix averages, and a non-universal
(i.e.~Hamiltonian-specific) part, which we also evaluate.
Comparing with Korepin's result leads to an explicit formula for
the central charge, which plays a fundamental role in the
conformal-field-theoretic approach.  In addition, we obtain the
sub-leading (constant) term in the asymptotics.

We also show that the random-matrix averages are related to
solutions of certain {\it Painlev\'e equations}.  In the case of
Hamiltonians possessing $\UN$ symmetry, these solutions, and thus
the entropy of the entanglement, can be determined exactly by
means of recurrence relations.

The structure of the paper is as follows.  In section~\ref{XYpres}
we review the results of Vidal {\it et al}~\cite{VLRK02} and Jin
and Korepin~\cite{JK03}.  In section~\ref{qfc} we introduce the
models that we investigate.  Sections~\ref{unitary},
\ref{orthogonal2n}, \ref{symplectic} and~\ref{ortodd} are devoted
to identifying the symmetry classes of Hamiltonians that are
associated to each compact group.  In section~\ref{Fisher-Hartwig}
we apply generalizations of the Fisher-Hartwig conjecture to
compute the entanglement for each symmetry class. Finally, in
section~\ref{Painleve} we investigate the connection between the
interpretation of the entanglement as an average over a compact
group and the theory of Painlev\'e equations.

\section{Bipartite entanglement in the XY model}
\label{XYpres}

We begin by reviewing the results of Vidal {\it et
al}~\cite{VLRK02} and Jin and Korepin~~\cite{JK03} concerning the
entanglement in the ground states of the XY and XX models. These
systems are spin-$1/2$ ferromagnetic chains with an exchange
coupling $\alpha$ in a constant transversal magnetic field $h$.
The Hamiltonian $H=hH_\alpha$ with $H_{\alpha}$ given by
\begin{equation}
\label{XYmodel}
H_{\alpha} = -\frac{\alpha}{2}\sum_{j = 0}^{M-1}
\left[(1 + \gamma)\sigma_j^x \sigma_{j+1}^x + (1-\gamma)\sigma_j^y
\sigma_{j+1}^y\right] - \sum_{j=0}^{M-1} \sigma_j^z,
\end{equation}
where $\sigma^a$ denotes the Pauli matrices and $a=x,y,z$. It is
sometimes convenient to work with the spin operators $S^a_j=
\sigma^a_j/2$ instead of the Pauli matrices. The parameter
$\gamma$ lies in the interval $[0,1]$ and measures the geometric
anisotropy of $H_{\alpha}$. Throughout this paper we will assume
periodic boundary conditions, i.e. $\sigma^a_M = \sigma^a_{0}$. At
zero temperature the system is in the ground state $\gsk$.  In the
limit $M \rightarrow \infty$ the Hamiltonian~\eqref{XYmodel}
undergoes a quantum phase transition at $\alpha_c = 1/(2h)$: the
spin-spin correlation functions $\gsb S^x_jS^x_k\gsk$ and $\gsb
S^y_jS^y_k \gsk $ decay algebraically at the critical point and
exponentially when $\alpha \neq \alpha_c$.

Let us partition the  Hamiltonian~\eqref{XYmodel} into two
subchains, which we denote by P and Q. The subsystem P is composed
of the first $N$ spins, with $1 \ll N \ll M$. The Hilbert space
where the Hamiltonian~\eqref{XYmodel} acts is the direct product
$\mathcal{H} = \mathcal{H}_{\rm P} \otimes \mathcal{H}_{{\rm Q}}$,
where $\mathcal{H}_{{\rm P}}$ and $\mathcal{H}_{{\rm Q}}$ are
generated by the spins in P and Q respectively and are spanned by
the vectors
\begin{equation}
\label{bvec} \prod_{j=0}^{N-1}(\sigma^{-}_j)^{r_j}
\ket{\boldsymbol{\Psi}_{{\rm F}}} \quad {\rm and } \quad
\prod_{j=N}^{M-1}(\sigma^{-}_j)^{r_j} \ket{\boldsymbol{\Psi}_{{\rm
F}}}, \quad r_j =0,1,
\end{equation}
where $\sigma^{\pm}_j = (\sigma^x_j \pm \rmi \sigma^y_j)/2$ and
$\ket{\boldsymbol{\Psi}_{{\rm F}}}$ denotes the ferromagnetic
state with all spins up. Since the ground state is non-degenerate
this subdivision creates a pure bipartite system. The measure of
entanglement is defined as the von Neumann entropy of either
subchain:
\begin{equation}
\label{vnent}
E_{{\rm P}}=E_{{\rm Q}} = -\trace \rho_{{\rm
P}}\log_2 \rho_{{\rm P}} = - \trace \rho_{{\rm Q}} \log_2
\rho_{{\rm Q}},
\end{equation}
where $\rho_{{\rm P}}$ and $\rho_{{\rm Q}}$ are the reduced
density matrices of P and Q, i.e.
\begin{equation}
\rho_{{\rm P}} = \trace_{{\rm Q}} \rho_{{\rm PQ}} \quad {\rm and}
\quad \rho_{{\rm Q}} = \trace_{{\rm P}} \rho_{{\rm PQ}}.
\end{equation}
The operators $\rho_{{\rm PQ}}$ form the density matrix of the
whole system, $\rho_{{\rm P Q}} = \gsk\negthinspace\gsb$.

Let us introduced the Jordan-Wigner transformations at each site
of the lattice $\{1,\ldots,M\}$:
\begin{subequations}
\begin{equation}
\label{mop}
m_{2l + 1} = \left(\prod_{j =0}^{l-1} \sigma_j^z
\right)\sigma_l^x \quad {\rm and} \quad m_{2l} =
\left(\prod_{j=0}^{l-1} \sigma_j^z\right)\sigma_l^y.
\end{equation}
The inverse relations are
\begin{equation}
\label{invrel}
\begin{split}
\sigma^z_l & = \rmi m_{2l} m_{2l + 1}, \\
\sigma_l^x & = \left(\prod_{j=0}^{l-1}\rmi
m_{2j}m_{2j+1}\right)m_{2l+1}, \\
\sigma_l^y &=
\left(\prod_{j=0}^{l-1}\rmi m_{2j}m_{2j+1}\right)m_{2l}
\end{split}
\end{equation}
\end{subequations}
These operators are Hermitian and obey the anticommutation
relations $\{m_j,m_k\}=2\delta_{jk}$.  We also define
\begin{equation}
\label{defermi} b_l = (m_{2l + 1} -\rmi m_{2l})/2 \quad \text{and}
\quad b_l^\dagger= (m_{2l + 1} +\rmi m_{2l})/2,
\end{equation}
which obey the anticommutation relations
\begin{equation}
\{b_j,b_k\} = 0 \quad \text{and} \quad \{b_j^\dagger,b_k\} =\delta_{jk}
\end{equation}
and therefore are Fermi operators.

The expectation values of
the operators~\eqref{mop} with respect to $\gsk$ are
\begin{subequations}
\label{exval}
\begin{gather}
\label{exval1}
\gsb m_k \gsk = 0, \\
\label{exval2}
\gsb m_jm_k \gsk  = \delta_{jk} + \rmi (C_M)_{jk},
\end{gather}
\end{subequations}
where the correlation matrix $C_M$ has the block structure
\begin{subequations}
\label{corrmat}
\begin{equation}
C_M = \begin{pmatrix} C_{11} & C_{12} & \cdots & C_{1M} \\
                      C_{21} & C_{22} & \cdots & C_{2M} \\
                      \hdotsfor[2]{4} \\
                      C_{M1} & C_{M2} & \cdots & C_{MM}
\end{pmatrix}
\end{equation}
with
\begin{equation}
\label{blockC}
C_{jk} = \begin{pmatrix} 0 & g_{j-k} \\
                         -g_{k-j} & 0
         \end{pmatrix}.
\end{equation}
\end{subequations}
For large $M$, the real numbers $g_l$ are the Fourier coefficients
\begin{equation}
\label{Fcoefa}
g_l = \frac{1}{2\pi}\int_0^{2\pi} \frac{\alpha \cos
\theta - 1 +\rmi \alpha\gamma \sin \theta}{\sqrt{\left(\alpha \cos
\theta - 1 \right)^2 + \gamma^2\alpha^2\sin^2
\theta}}\,\rme^{-\rmi l \theta} \rmd \theta.
\end{equation}
Formulae~\eqref{exval} were first computed (in a slightly
different form) by Lieb {\it et al}~\cite{LSM61} and Barouch and
McCoy~\cite{BM71}. The expectation values of products of arbitrary
numbers of the operators~\eqref{mop} can be obtained
from~\eqref{exval} using Wick's theorem. Equation~\eqref{exval1}
follows from the invariance of $H_{\alpha}$ under the map
$(\sigma_j^x,\sigma_j^y,\sigma_j^z)\mapsto
(-\sigma_j^x,-\sigma_j^y,\sigma^z_j)$. In appendix~B we
extend~\eqref{exval2} to a class of Hamiltonians that generalizes
the XY model.

In order to compute the entropy~\eqref{vnent} we need an
expression for the reduced density matrix $\rho_{{\rm P}}$.  We
therefore restrict our analysis to the first $N$ consecutive
spins. The correlation matrix is simply constructed by removing
the last $2(M - N)$ rows and columns from the
matrix~\eqref{exval2}; we shall denote this restriction by $C_N$.
If $V$ is an orthogonal matrix, then the operators
\begin{equation}
\label{dop}
d_j = \sum_{k=0}^{2N-1} V_{jk}m_k
\end{equation}
are Hermitian and obey the anti-commutation relations $\{d_j,d_k\}
= 2\delta_{jk}$ too. Furthermore, since the Fourier coefficients
defined in equation~\eqref{Fcoefa} are real, there exists an
appropriate $V \in {\rm SO}(2N)$ that block-diagonalizes $C_N$:
\begin{equation}
\label{bdiag} V C_N V^t
= \bigoplus_{j=0}^{N-1} \nu_j
\begin{pmatrix} 0 & 1
\\-1 & 0\end{pmatrix},
\end{equation}
where the $\nu_j$ are real and, for reasons that will become
apparent later, lie in the interval $[-1,1]$.  Now, let us
define
\begin{equation}
\label{cfermop} c_j = (d_{2j+1} - \rmi d_{2j})/2, \quad
j=0,\ldots, N-1,
\end{equation}
which are Fermi operators, i.e.
\begin{equation}
\{c_j,c_k\}=0 \quad {\rm and} \quad
\{c^{\dagger}_j,c_k\}=\delta_{jk}
\end{equation}
and are linearly related to those ones defined in~\eqref{defermi}. A basis for
$\mathcal{H}_{{\rm P}}$ is also given by the $2^N$ vectors
\begin{equation}
\prod_{j=0}^{N-1}(c^\dagger_j )^{r_j} \ket{\boldsymbol{\Psi}_{{\rm
vac}}}, \quad r_j=0,1,
\end{equation}
where the state $\ket{\boldsymbol{\Psi}_{{\rm vac}}}$ is defined
by the condition
\begin{equation}
 c_j\ket{\boldsymbol{\Psi}_{{\rm vac}}} = 0, \quad
 j=0,\ldots,N-1.
\end{equation}
The expectation values of the $c_j$s are readily obtained from
equation~\eqref{bdiag}:
\begin{subequations}
\label{exvalfe}
\begin{gather}
\gsb c_j\gsk = \gsb c_j\, c_k \gsk = 0, \\
\gsb c_j^\dagger \,c_k\gsk = \delta_{jk}\frac{1 - \nu_j}{2}.
\end{gather}
\end{subequations}

The reduced density matrix $\rho_{{\rm P}}$ can be computed
directly from the expectation values~\eqref{exvalfe}; we describe
the details of the derivation in appendix~A. The final expression
is
\begin{equation}
\label{rop} \rho_{{\rm P}} = \prod_{j=0}^{N-1}\left(\frac{1
-\nu_j}{2}\,c^\dagger_j \, c_j + \frac{1 + \nu_j}{2}\, c_j \,
c^\dagger_j \right).
\end{equation}
In other words, as equations~\eqref{exvalfe} already suggest,
these fermionic modes are in a product of uncorrelated states,
therefore the density matrix is the direct product
\begin{equation}
\rho_{{\rm P}} = \bigotimes_{j=0}^{N-1} \rho_j \quad {\rm with}
\quad \rho_j = \frac{1 - \nu_j}{2}\,c^\dagger_j \, c_j + \frac{1 +
\nu_j}{2}\, c_j \, c^\dagger_j.
\end{equation}
As a consequence, if one of the $\nu_j$s lies outside the interval
$[-1,1]$, then either $(1 + \nu_j)/2$ or $(1-\nu_j)/2$ would be
negative and so $\rho_j$ could not be a density matrix.  At this
point the entropy of the entanglement between the two subsystems
can be easily derived from equation~\eqref{vnent}:
\begin{equation}
\label{nent}
E_{{\rm P}} = \sum_{j=0}^{N -1} \rme(1,\nu_j),
\end{equation}
where
\begin{equation}
\label{binaryent}
\rme(x,\nu)= - \frac{x +
\nu}{2}\log_2\left(\frac{x + \nu}{2}\right) - \frac{x -
\nu}{2}\log_2\left(\frac{x - \nu}{2}\right).
\end{equation}

Vidal {\it et al}~\cite{VLRK02} computed $E_{{\rm P}}$ numerically
for the XY model and observed that for values of $\alpha$ and
$\gamma$ close to a critical point it grows logarithmically with
$N$, while away from the phase transitions it is either zero or
saturates to a constant value. Jin and Korepin~\cite{JK03}
computed $E_{{\rm P}}$ when $\gamma =0$, i.e. for the XX model,
using the Fisher-Hartwig conjecture, which gives the leading order
asymptotics of determinants of Toeplitz matrices whose symbols
have zeros or discontinuities. This is also central to our own
approach, so we will give a brief account of their method.

When $\gamma = 0$ the numbers~\eqref{Fcoefa} are the Fourier
coefficients of the step function
\begin{equation}
g(\theta) = \begin{cases} 1 & \text{if $-k \le \theta < k$} \\
                         -1 & \text{if $ k \le \theta < 2\pi
                         -k$},
            \end{cases}
\end{equation}
where $k = \arccos (1/\alpha)$. In order to be in a critical
regime, the parameter $1/\alpha$ must lie in the interval
$(-1,1)$; for $\abs{\alpha} \le 1$ the entanglement $E_{{\rm P}}$
is trivially zero.  In physical terms this means that outside the
critical regime the magnetic field is strong enough to align all
the spins, therefore entangled states cannot appear.  Since
$g(\theta)$ is even its Fourier coefficients have the symmetry
$g_l = g_{-l}$, therefore the correlation matrix $C_N$ factorizes
into the direct product
\begin{equation}
\label{toepf} C_N = T_N[g] \otimes \begin{pmatrix} 0 & 1 \\ -1 & 0
\end{pmatrix},
\end{equation}
where $T_N[g]$ is the matrix
\begin{equation}
(T_N[g])_{jk} = g_{j-k}.
\end{equation}
 Matrices of this type are called {\it Toeplitz matrices} and
  $g(\theta)$ is called the {\it symbol} of $T_N[g]$.
 Toeplitz matrices and their determinants, known as {\it Toeplitz
 determinants}, will play an important role in our analysis.

 As a consequence of~\eqref{toepf} the
$\nu_j$s are simply the eigenvalues of $T_N[g]$. One can then
rewrite~\eqref{nent} using Cauchy's theorem~\cite{JK03}:
\begin{equation}
\label{JKidea}
E_{{\rm P}} = \lim_{\epsilon \rightarrow 0^+}
\lim_{\delta \rightarrow 0^+} \frac{1}{2\pi \rmi}
\oint_{c(\epsilon,\delta)} \rme(1 + \epsilon,\lambda) \frac{\rmd
\ln D_N[g](\lambda)}{\rmd \lambda} \rmd \lambda,
\end{equation}
where $D_N(\lambda)$ is the characteristic polynomial
\begin{equation}
D_N[g](\lambda) = \det\left(\lambda I - T_N[g]\right),
\end{equation}
which is also a Toeplitz determinant with symbol
\begin{equation}
g(\theta;\lambda) = \begin{cases} \lambda - 1 & \text{if $-k \le \theta < k$} \\
                         \lambda + 1 & \text{if $ k \le \theta < 2\pi
                         -k$}.
            \end{cases}
\end{equation}
The contour of integration $c(\epsilon,\delta)$ depends on the
parameters $\epsilon$ and $\delta$ and includes the interval
$[-1,1]$; as $\epsilon$ and $\delta$ tend to zero the contour
approaches the interval $[-1,1]$.  This technical expedient is
necessary in order to guarantee that the branch points of
$\rme(1 + \epsilon,\lambda)$ lie outside the contour of integration
and thus that $\rme(1 + \epsilon,\lambda)$ is analytic inside
$c(\epsilon,\delta)$.
As already mentioned, the polynomial
$D_N(\lambda)$ can be estimated for large $N$ using a formula,
which we will discuss in detail in section~\ref{Fisher-Hartwig},
known as the {\it Fisher-Hartwig conjecture}~\cite{FH68}. This
asymptotic formula can then be inserted in equation~\eqref{JKidea}
and the integral explicitly evaluated. We will not go into the
details of the computation, which can be found in Jin and
Korepin's original paper. Their result is that for the XX model
\begin{equation}
E_{{\rm P}} \sim \frac{1}{3}\log_2 N, \quad N \rightarrow \infty.
\end{equation}
This logarithmic divergence is a direct consequence of a general
theorem on the spectra of Toeplitz matrices~\cite{BM94}, which in
our case reduces to the statement that in the limit $N \rightarrow
\infty$ all except for ${\rm O}(\ln N)$ of the eigenvalues of
$T_N[g]$ migrate toward the points $1$ and $-1$, where, trivially,
we have $\rme(1,1)=\rme(1,-1)=0$.

We now make an observation relating to~\eqref{JKidea} that will
have important consequences for the way we shall proceed.
Toeplitz determinants are equivalent to averages of appropriate
functions over the group of unitary matrices $\UN$. Let us make
this statement more precise. The group $\UN$ is compact and
therefore has a unique left and right invariant measure known as
Haar measure. The invariance of the measure makes it a natural
probability density for unitary matrices, because different
regions in $\UN$ are equally weighted. This is usually referred to
as the {\it Circular Unitary Ensemble} (CUE) of random matrices.
After having integrated over the degrees of freedom associated
with the eigenvectors, the explicit expression of Haar measure
becomes
\begin{equation}
P_{\UN}(\theta_1,\ldots,\theta_N) = \frac{1}{(2\pi)^N
N!}\prod_{1\le j < k \le N}\abs{\rme^{\rmi \theta_j} - \rme^{\rmi
\theta_k}}^2.
\end{equation}
Now, let $G(U)$ be a function on $\UN$ that depends only on the
eigenvalues of $U$ and is such that
\begin{equation}
G(U) = \prod_{j=1}^{N}g(\theta_j),
\end{equation}
where $g(\theta)$ is $2\pi$-periodic.  An identity due to
Heine~\cite{Hei78} and --- in the form that we use ---
Szeg\H{o}~\cite{Seg59}, pp. 27 and 288,  asserts that
\begin{equation}
\label{HSid}
\begin{split}
\Bigl \langle G(U) \Bigr \rangle_{{\UN}} &=\left \langle
\prod_{j=1}^N g(\theta_j)\right \rangle_{{\UN}}\\
& = \frac{1}{(2\pi)^N N!}\int_0^{2\pi}\cdots \int_0^{2\pi}
\left(\prod_{j=1}^N g(\theta_j)\right) \prod_{1\le j < k \le
N}\abs{\, \rme^{\rmi \theta_j} - \rme^{\rmi \theta_k}}^2\rmd
\theta_1\cdots \rmd \theta_N   \\
& = \det\left(g_{j-k}\right)_{j,k=0,\ldots,N-1},
\end{split}
\end{equation}
where
\begin{equation}
g_l = \frac{1}{2\pi}\int_0^{2\pi}g(\theta)\, \rme^{-\rmi l
\theta}\rmd \theta.
\end{equation}
Throughout this paper the brackets $\left \langle \cdot \right
\rangle$ denote the average with respect to Haar measure (not
necessarily just that of $\UN$).

For the XX model this remarkable identity allows us to
express~\eqref{JKidea}  as an average over $\UN$. Two questions
now arise.  First, can~\eqref{JKidea} be generalized to other
Hamiltonians? And second, since there are no obvious reasons why
$\UN$ should be singled out with respect to the other compact
groups, are there systems for which the symmetries of the
interaction give rise to a von Neumann entropy that can be
re-expressed in terms of averages over $\SpN$ and
$\mathrm{O}^\pm(N)$?

We note first that when $\gamma \neq 0$ the symbol of the matrix
$T_N[g]$ is
\begin{equation}
\label{defsyma} g(\theta)=\frac{\alpha \cos \theta - 1 +\rmi
\alpha\gamma \sin \theta}{\sqrt{\left(\alpha \cos \theta -1
\right)^2 + \gamma^2\alpha^2\sin^2 \theta}}.
\end{equation}
The numbers $\nu_j$ that appear in formula~\eqref{bdiag} are not
the eigenvalues of $T_N[g]$, which is not symmetric and as a
consequence need not have a real spectrum. Instead they are the
eigenvalues of the matrix $S = \left(T_N[g]T_N[g]^t\right)^{1/2}$.
As a consequence, the entanglement is not straightforwardly
expressible in terms of a Toeplitz determinant and so in terms of
an average over $\UN$.  Even so, it can again be expressed as an
integral transform of the characteristic polynomial of $S$.  It is
worth remarking that the spin-spin correlation functions can still
be expressed in terms of Toeplitz determinants and so related to
averages over $\UN$.

In the following sections we shall show that for Hamiltonians that
can be expressed as quadratic forms of fermionic oscillators, of
which the XY model is an example, there exists a one-to-one
correspondence between the symmetries of the Hamiltonian and the
interpretation of the von Neumannn entropy in terms of averages
over one the classical compact groups when $\gamma = 0$. In
section~\ref{Painleve} we establish a connection between a
generalization of formula~\eqref{JKidea} and the theory of
Painlev\'e equations.

\section{Quadratic chains of fermionic operators}
\label{qfc} The most
general form of Hamiltonian related to quantum spin chains is
\begin{equation}
\label{impH}
H_{\alpha} = \alpha \left[\sum_{j,k=0}^{M-1}
b^\dagger_jA_{jk} b_k + \frac{\gamma}{2}\left(b^\dagger_j
B_{jk}b^\dagger_k - b_j B_{jk}b_k \right)\right] -
2\sum_{j=0}^{M-1} b^{\dagger}_jb_j,
\end{equation}
where $\alpha$ and $\gamma$ are real parameters, $0\le \gamma
\le 1$ and the $b_j$s are the Fermi oscillators defined
in equation~\eqref{defermi}. We take periodic boundary conditions, i.e. $b_M = b_0$.
Since $H_{\alpha}$ is Hermitian, $A$ must be a Hermitian matrix,
and because of the anticommutation relation of the Fermi
operators, $B$ must be antisymmetric. Without loss of generality,
we will consider only matrices $A$ and $B$ with real elements.

Up to a overall constant, the XY model~\eqref{XYmodel} maps into
the Hamiltonian~\eqref{impH}\footnote{This is strictly true only
for open-end Hamiltonians.  If we assume periodic boundary
conditions, then the term $b^{\dagger}_{M-1}b_0$ should be
replaced by $\left[\prod_{j=0}^{M-1}\left(2b^\dagger_jb_j
-1\right)\right]b_{M-1}b_0$.  However, because we are interested
in the limit $M \rightarrow \infty$, the extra factor in front of
$b_{M-1}b_0$ can be neglected.} with the matrices $A$ and $B$
given by
\begin{equation}
\label{ABmat}
A   = \begin{pmatrix}
 0 & 1 & 0  &\cdots & 0 & 0 & 1 \\
                1& 0 & 1 &  \cdots &0 & 0 & 0 \\
                0 & 1 & 0 & \cdots & 0 & 0 & 0 \\
                                 \hdotsfor[2]{7} \\
                0 & 0 & 0  & \cdots & 1 & 0 & 1 \\
                1 & 0 & 0 & \cdots & 0 & 1 & 0
\end{pmatrix}
\quad {\rm and} \quad
B = \begin{pmatrix} 0 & 1 & 0 &\cdots & 0 & 0 & -1 \\
                -1& 0 & 1 & \cdots &0 & 0 & 0 \\
                0 & -1 & 0  & \cdots & 0 & 0 & 0 \\
                                 \hdotsfor[2]{7} \\
                0 & 0 & 0 & \cdots & -1 & 0 & 1 \\
                1 & 0 & 0 & \cdots & 0 & -1 & 0
      \end{pmatrix}.
\end{equation}
Similarly, equation~\eqref{invrel} maps the
Hamiltonian~\eqref{impH} into the spin chain
\begin{equation}
\label{genmodel2}
\begin{split}
H_{\alpha} & = -\frac{\alpha}{2} \sum_{0 \le j \le k \le M-1}
 \left[(A_{jk} + \gamma B_{jk})\sigma_j^x
\sigma_k^x \left(\prod_{l=j+1}^{k-1}\sigma_l^z\right) \right. \\
& \quad +(A_{jk}-\gamma B_{jk})\sigma_j^y \left.
\sigma_k^y\left(\prod_{l=j+1}^{k-1}\sigma_l^z\right) \right]
-\sum_{j=0}^{M-1}\sigma_j^z.
\end{split}
\end{equation}
In other words, the two systems~\eqref{impH} and~\eqref{genmodel2}
are equivalent.  One sees that the second sum in~\eqref{impH} is
the analogue of a uniform magnetic field, the parameter $\gamma$
introduces a geometric anisotropy in the interaction, and $\alpha$
is an exchange coupling constant.

We will here be concerned with the entanglement between the first
$N$ oscillators and the rest of the chain, when the system is in
the ground state $\gsk$ and as the length of chain tends to
infinity. In a similar fashion as for the XY model we decompose
the Hilbert space into the direct product
$\mathcal{H}=\mathcal{H}_{{\rm P}} \otimes \mathcal{H}_{{\rm Q}}$,
where $\mathcal{H}_{{\rm P}}$ is generated by the first $N$
sequential oscillators and $\mathcal{H}_{{\rm Q}}$ by the
remaining $M - N$.   They are spanned by the vectors
\begin{equation}
\prod_{j=0}^{N-1}(b^\dagger_j )^{r_j} \ket{\boldsymbol{\Psi}_{{\rm
vac}}} \quad {\rm and } \quad \prod_{j=N}^{M-N}(b^\dagger_j
)^{r_j} \ket{\boldsymbol{\Psi}_{{\rm vac}}}, \quad r_j=0,1,
\end{equation}
respectively, where the vacuum state $\ket{\boldsymbol{\Psi}_{{\rm
vac}}}$ is defined by
\begin{equation}
 b_j\ket{\boldsymbol{\Psi}_{{\rm vac}}} = 0, \quad
 j=0,\ldots,M-1.
\end{equation}
Our goal is to determine the asymptotic behaviour for large $N$,
with $N = {\rm o}(M)$, of the von Neumann entropy
\begin{equation}
\label{entang2}
E_{{\rm P}} = - \trace \rho_{{\rm P}} \log_2
\rho_{{\rm P}},
\end{equation}
where $\rho_{{\rm P}}= \trace_{{\rm Q}} \rho_{{\rm PQ}}$ and
$\rho_{{\rm PQ}}= \gsk\negthinspace\gsb$.

 The first step involves determining the expectation values with respect to $\gsk$ of
products of arbitrary numbers of operators~\eqref{mop}.  From the
invariance of the Hamiltonian~\eqref{impH} under the
transformation $b_j \mapsto -b_j$, it follows that $\gsb m_l
\gsk=0$; for the same reason, the expectation value of the product
of an odd number of $m_j$s must be zero.  The expectation values
of the product of an even number of $m_j$s can be computed using
Wick's theorem
\begin{equation}
\label{Wick-Th}
\gsb m_{j_1}m_{j_2}\cdots m_{j_{2n}} \gsk =
\sum_{\text{ all pairings}} (-1)^p \prod_{\text{all pairs}}
\left(\text{contraction of the pair}\right),
\end{equation}
where a contraction of a pair is defined by $\gsb m_{j_l}m_{j_m}
\gsk$ and $p$ is the signature of the permutation, for a given
pairing, necessary to bring operators of the same pair next to one
other from the original order.

Before continuing our analysis it is worth noting that the
spin-spin correlation functions $\gsb S^x_jS^x_k\gsk$ and $\gsb
S^y_jS^y_k \gsk$ of the XY models are products of the
type~\eqref{Wick-Th}. These correlations were first studied by
Barouch and McCoy~\cite{BM71}, who showed that in general they are
Pfaffians, which, when the system reaches thermal equilibrium,
reduce to Toeplitz determinants and can be computed using
Szeg\H{o}'s theorem and the Fisher-Hartwig conjecture. The
different behaviour of these determinants away from and in the
proximity of critical points determines how quantum phase
transitions appear. Thus, in the same way that we can ask which
features of the Hamiltonian~\eqref{impH} lead to expressions for
the entropy of entanglement in terms of an average over $\UN$, so
we can enquire which properties of $H_{\alpha}$ ensure that its
physical correlations are expressible as averages over $\UN$. In
view of the way physical correlations and entanglement affect each
other, these two questions are closely related. Similarly, we can
pose the same questions for the other classical compact groups
$\SpN$ and $\mathrm{O}^\pm(N)$.

The expectation values $\gsb m_j m_k \gsk$ can be deduced from the
work of Lieb {\it et al}~\cite{LSM61}; as already mentioned in
section~\ref{XYpres}, we give the details of this computation in
appendix~B.  They generalize formula~\eqref{exval2}, in that the
Fourier coefficients $g_{j-k}$ and $g_{k-j}$ in the $2\times 2$
block~\eqref{blockC} are replaced by the matrix elements
$(T_M)_{jk}$ and $(T_M)_{kj}$, where the matrix $T_M$ is defined
by
\begin{equation}
\label{Tmatr}
\left(T_M\right)_{jk} = \sum_{l=0}^{M-1}\psi_{l
j}\phi_{lk}, \quad j,k=0,\ldots, M-1,
\end{equation}
and the vectors $\vphi_k$ and $\vpsi_k$ are real and orthogonal
and obey the eigenvalue equations
\begin{subequations}
\label{stepb}
\begin{align}
\label{phi2b}
\alpha^2(A -(2/\alpha)I-\gamma B)(A -(2/\alpha)I
+\gamma B)\vphi_k & =
\abs{\Lambda_k}^2\vphi_k, \\
\label{psi2b} \alpha^2(A -(2/\alpha)I+\gamma B)(A
-(2/\alpha)I-\gamma B)\vpsi_k & = \abs{\Lambda_k}^2\vpsi_k.
\end{align}
\end{subequations}
These vectors are related by
\begin{subequations}
\label{fstepb}
\begin{align}
\label{phi1b}
\alpha(A -(2/\alpha)I + \gamma B)\vphi_k &= \abs{\Lambda_k} \vpsi_k, \\
\label{psi1b} \alpha(A - (2/\alpha)I - \gamma B)\vpsi_k &=
\abs{\Lambda_k}\vphi_k.
\end{align}
\end{subequations}
From equations~\eqref{stepb} it follows that the $\vphi_k$s and
$\vpsi_k$s are eigenvectors of positive symmetric matrices,
therefore the corresponding eigenvalues are always real and
positive; for later convenience, here and in appendix~B we express
them as the square of the absolute value of a complex number.

Now, the derivation of the von Neumann entropy~\eqref{entang2} is
identical to the analogous computation for the XY model described
in section~\ref{XYpres}. The formula for the entropy of the
subchain P that one obtains is
\begin{equation}
\label{JKidea2}
E_{{\rm P}} = \lim_{\epsilon \rightarrow 0^+}
\lim_{\delta \rightarrow 0^+} \frac{1}{2\pi \rmi}
\oint_{c(\epsilon,\delta)} \rme(1 + \epsilon,\lambda) \frac{\rmd
\ln D_N(\lambda)}{\rmd \lambda} \rmd \lambda,
\end{equation}
where
\begin{equation}
\label{chpoln}
 D_N(\lambda) = \det\left(I\lambda - S \right),
\end{equation}
$S$ is the real symmetric matrix $\left(T_N T_N^t\right)^{1/2}$
and $T_N$ is obtained from the matrix~\eqref{Tmatr} by removing
the last $M-N$ rows and columns. The path of integration of the
integral in~\eqref{JKidea2} is the same as that in~\eqref{JKidea}.
In order for~\eqref{JKidea2} to have physical meaning, the
eigenvalues of $S$ must all be in the interval $[-1,1]$.  This
will have to be verified for the various Hamiltonians of the
type~\eqref{impH} that we shall consider.

\section{$\UN$ symmetry}
\label{unitary}

If there is a connection between the entanglement of the ground
state or, more generally, between the physical correlations of the
ground state and the classical compact groups, it must be
reflected somehow in the symmetries of the system of fermionic
oscillators~\eqref{impH}.  The most obvious symmetry that the XY
model has in common with $\UN$ is the invariance under
translations: for the XY model it manifests itself in the fact
that the matrices~\eqref{ABmat} are mapped into themselves if
their rows and columns are simultaneously shifted by the same
integer $q$ (the periodicity of $H_{\alpha}$ is of course
inherited by the periodicity of the rows and columns of $A$ and
$B$); for $\UN$ such symmetry is reflected in the invariance of
the integral~\eqref{HSid} if all the $\theta_j$s are shifted by
the same amount $\chi$.   It is worth noting that the
integral~\eqref{HSid} can also be written in the form
\begin{equation}
\label{altfo}
\left \langle \prod_{j=1}^N g(\theta_j)\right
\rangle_{\UN}=\int_0^{2\pi}\cdots \int_0^{2\pi}
\left(\prod_{j=1}^N g(\theta_j)\right) \det\left[S_N\left(\theta_j
- \theta_k\right)\right]_{j,k=1,\ldots,N}\rmd \theta_1\cdots \rmd
\theta_N,
\end{equation}
where $S_N(z)$ is the kernel of the Haar measure of $\UN$ and is
given by
\begin{equation}
\label{UNkernel}
S_N(z) =
\frac{1}{2\pi}\frac{\sin(Nz/2)}{\sin(z/2)}.
\end{equation}
The invariance under translations of the Haar measure is then
evident in the kernel~\eqref{UNkernel}. We shall now determine a
general formula for the matrix~\eqref{Tmatr} in the limit $M
\rightarrow \infty$ by assuming that the physical interaction in
the Hamiltonian~\eqref{impH} is invariant under translations; we
shall also deduce under which conditions the characteristic
polynomial appearing in the integrand of formula~\eqref{JKidea2}
can be interpreted as an average over $\UN$.

In order to simplify the algebra we shall denote $\bA =\alpha A -
2I$ and $\bB = \alpha \gamma B$. If $H_{\alpha}$ is invariant
under translations of the lattice $\{0,1,\ldots,M-1\}$, than the
elements of the matrices $\bA$ and $\bB$ must depend only on the
difference between the row and column indices, i.e. $\bA$ and
$\bB$ must be Toeplitz matrices.  In addition, because of the
periodic boundary conditions, $\bA$ and $\bB$ must be {\it
cyclic}.

Now, let $a$ and $b$ be two real functions on
$\mathbb{Z}/M\mathbb{Z}$, even and odd respectively. The matrix
elements of $\bA$ and $\bB$ can be written as
\begin{equation}
\label{circdef}
\bA_{jk}=a(j-k) \quad {\rm and} \quad
\bB_{jk}=b(j-k),\quad j,k=0,\ldots,M-1.
\end{equation}
A complete set of orthonormal eigenvectors of cyclic matrices are
the complex exponentials:
\begin{equation}
\label{ceigAB} \phi_{kj}= \frac{\exp\left(\frac{2\pi \rmi k
j}{M}\right)}{\sqrt{M}}, \quad j,k=0,\ldots, M-1,
\end{equation}
as can be easily verified by direct substitution. Now, the
matrices $\bA$ and $\bB$ defined in equation~\eqref{circdef}
commute. This becomes straightforward if we notice that
$(\bA\bB)_{jk}$ is the convolution of $a$ and $b$ evaluated at
$j-k$:
\begin{equation}
\begin{split}
(\bA\bB)_{jk} &= \sum_{l =0}^{M-1} a(j-l)b(l-k) = \sum_{l=0}^{M-1}
a(l)b(j-k -l) \\
& =\sum_{l=0}^{M-1} a(l-k)b(j-l) = (\bB\bA)_{jk}.
\end{split}
\end{equation}
As a consequence, the complex exponentials~\eqref{ceigAB} are a
complete set of eigenvectors of $\bA + \bB$ too. Therefore, we
have
\begin{equation}
\label{expleq}
\sum_{l=0}^{M-1}\left[a(p-l) +
b(p-l)\right]\phi_{kl}= \Lambda_k' \phi_{kp} =
\abs{\Lambda_k}\psi_{kp},
\end{equation}
with $\vpsi_{k}= \vphi_{k} \Lambda'_{k}/\abs{\Lambda_k}$. Because
both $\vphi_k$ and $\vpsi_k$ are normalized,
$\Lambda_k'/\abs{\Lambda_k}$ is a phase factor and we can set
$\Lambda_k'=\Lambda_k$.  It is important to notice that since we
have taken the $\vphi_k$s and $\vpsi_k$s to be common eigenvectors
of $A$ and $B$ they are not in general real. The
matrix~\eqref{Tmatr} should therefore be replaced by
\begin{equation}
\label{Tmatra} \left(\overline T_M\right)_{jk} =
\sum_{l=0}^{M-1}\overline \psi_{lj} \phi_{lk}, \quad j,k=0,\ldots,
M-1,
\end{equation}
which a priori need not be real.  However, the
matrix~\eqref{Tmatra} is unitarily equivalent to~\eqref{Tmatr};
indeed, we shall prove that~\eqref{Tmatra} is real too and as
consequence there is an orthogonal matrix $O$ such that
$OT_MO^t=\overline T_M$. Therefore, we shall not distinguish
between them.

The eigenvalues of $\bA + \bB$ can be determined simply by
inserting the eigenvectors~\eqref{ceigAB} into the left-hand-side
of equation~\eqref{expleq} and using the parities of the functions
$a(j)$ and $b(j)$. We have
\begin{equation}
\label{fexpu} \Lambda_k = \begin{cases} a(0) + 2\sum_{j =
1}^{(M-1)/2}\left(a(j)\cos
kj + \rmi b(j)\sin kj\right) & \text{if $M$ is odd} \\
 a(0) + (-1)^l a(M/2)  + 2\sum_{j = 1}^{M/2-1}\left(a(j)\cos
kj  + \rmi b(j) \sin kj\right)& \text{if $M$ is even,}
\end{cases}
\end{equation}
where $k$ does not denote an integer anymore but the wave number
\begin{equation}
k = \frac{2\pi l}{M}, \quad l=0,\ldots,M-1.
\end{equation}
 The matrix~\eqref{Tmatra} now becomes
\begin{equation}
\label{vgcu} (T_M)_{jl}=  \frac{1}{2\pi}\sum_{k=0}^{2\pi\left(1 -
1/M\right)} \frac{\overline \Lambda_k}{\abs{\Lambda_k}}
 \rme^{-\rmi k (j -l)} \Delta k,
\end{equation}
where $\Delta k = 2\pi/M$.  From~\eqref{vgcu} it follows that
taking the complex conjugate of the right-hand side
of~\eqref{vgcu} is equivalent to replacing $k$ by $2\pi -k$ in the
sum, therefore $T_M$ is real. If now we focus on the system
consisting of the first $N$ oscillators and let $M \rightarrow
\infty$, we obtain
\begin{equation}
\label{Toeplitzr} (T_N)_{jk} \xrightarrow[M \rightarrow \infty]{}
 \frac{1}{2\pi}\int_0^{2\pi}\frac{\Lambda(\theta)}{\abs{\Lambda(\theta)}}
\rme^{-\rmi \left(j - k\right)\theta}\rmd \theta,
\end{equation}
where $\Lambda(\theta)$ is the periodic function
\begin{equation}
\label{fLam}
\Lambda(\theta)=\sum_{j=-\infty}^\infty \Lambda_j \,
\rme^{\rmi j \theta}
\end{equation}
with $\Lambda_j=a(j)-b(j)$ if $j>0$ and $\Lambda_j=a(j)+b(j)$ if
$j<0$.  (We have also implicitly assumed that as $j \rightarrow
\infty$, $a(j)$ and $b(j)$ tend to zero sufficiently fast for the
Fourier series~\eqref{fLam} to converge.) Thus, $T_N[g]$ is a
Toeplitz matrix with symbol $g(\theta)=
\Lambda(\theta)/\abs{\Lambda(\theta)}$. It is worth emphasizing
that formula~\eqref{Toeplitzr} has been obtained by assuming only
the translation invariance of the Hamiltonian~\eqref{impH} and
periodic boundary conditions. Finally, if we define
\begin{equation}
\delta_l(j) = \begin{cases} 1 & \text{if $l \equiv j \bmod N$} \\
                           0 & \text{otherwise,}
 \end{cases}
\end{equation}
then we recover the $XY$ model with  the choice
\begin{equation}
\label{aforXYmod}
a(j) = \alpha\left[\delta_1(j) +
\delta_1(-j)\right] -2 \delta_0(j) \quad {\rm and} \quad
b(j)=-\alpha \gamma \left[ \delta_1(j) - \delta_1(-j)\right].
\end{equation}

At this point a few remarks relating to~\eqref{Toeplitzr} should
be made. First, because the interaction is invariant under
translations, we can take any set of $N$ consecutive oscillators
and the corresponding matrix $T_N[g]$ will still be a Toeplitz
matrix. Second, as a consequence of the Jordan-Wigner
transformations~\eqref{invrel} and Wick's theorem~\eqref{Wick-Th}
the spin-spin correlation functions are Toeplitz detrminants, i.e.
averages over $\UN$. Finally, there are important implications for
formula~\eqref{JKidea2}.  We mentioned that in order to have
physical meaning the eigenvalues of $(T_N[g]T_N[g]^t)^{1/2}$ must
lie in the interval $[-1,1]$.  Since the symbol $g(\theta)$ has
absolute value one, a theorem on the spectrum of Toeplitz
matrices~\cite{BM94} --- the same theorem mentioned in
section~\ref{XYpres} --- states that the eigenvalues of $T_N[g]$
are all inside the unit circle and approach the image of $g$ in
the limit $N \rightarrow \infty$.  It follows  that all the
eigenvalues of $(T_N[g]T_N[g]^t)^{1/2}$ lie in the interval
$[-1,1]$. It remains to establish when~\eqref{JKidea2} is an
average over $\UN$. The condition is that $T_N[g]$ should be
symmetric, in which case the correlation matrix $C_N$ factorizes
into the direct product as in~\eqref{toepf}. A necessary and
sufficient condition in order for $T_N[g]$ to be symmetric is that
$\Lambda(\theta)$ should be real and even, or equivalently
$\gamma$ should be zero; in other words, the interaction in the
Hamiltonian~\eqref{impH} must be isotropic. When $\gamma =0$ the
symbol $g(\theta)$ is a piece-wise continuous function that takes
the values $1$ and $-1$ and has discontinuities at all points
$\theta_r$ where the equation
\begin{equation}
\label{jloc}
 \Lambda(\theta_r) = 0
\end{equation}
is satisfied, with the additional condition that the first
non-zero derivative of $\Lambda(\theta)$ at $\theta_r$ is odd. 

\section{$\ONep$ symmetry}
\label{orthogonal2n}

We now address the question of finding a class of symmetries of
the Hamiltonian~\eqref{impH} which leads to an interpretation of
the spin-spin correlation functions and the formula for the
entropy of the entanglement~\eqref{JKidea2} as averages over
$\ONep$, the group of orthogonal matrices of dimension $2N\times
2N$ and determinant $1$.  We have seen that the expression of the
von Neumann entropy in terms of an average over $\UN$ is a direct
a consequence of the invariance under translations of the
Hamiltonian~\eqref{impH} and of its geometrical isotropy. We now
proceed in the same way as with $\UN$ and try to infer how the
structure of the kernel of the Haar measure of $\ONep$ is
reflected into the invariance properties of $H_{\alpha}$.

Eigenvalues of orthogonal and symplectic matrices come in complex
conjugate pairs, therefore $\ONep$ has only $N$ independent
eigenvalues. When dealing with the classical compact groups, we
shall adopt the convention of denoting by $\tilde{N}$ the total
number of eigenvalues and by $N$ the number of independent ones.
In general we shall denote an arbitrary group by $\GM$. Each of
the classical compact groups is identified by specific values of
$(\sigma_1,\sigma_2)$.  This correspondence is described in
appendix~C; for $\ONep$, $(\sigma_1,\sigma_2)=(1/2,1/2)$. Let
$F(U)$ be a {\it class function} on $\GM$, i.e. a symmetric
function depending only on the eigenvalues of $U$. Furthermore,
suppose that
\begin{equation}
F(U) = \prod_{j=1}^{\tilde N} f(\theta_j),
\end{equation}
where $f(\theta)$ is even and $2\pi$-periodic. The averages
discussed in appendix~C can all be written as
\begin{equation}
\label{altfo2}
\begin{split}
\Bigl \langle F(U)\Bigr \rangle_{\GM} & =\left \langle
\prod_{j=1}^{\tilde N} f(\theta_j)\right \rangle_{\GM}
\\
& = \int_{-\pi}^{\pi}\cdots \int_{-\pi}^{\pi} \det
\left[f(\theta_j)f(\theta_k)Q_N^{(\sigma_1,\sigma_2)}
\left(\theta_j,\theta_k\right)\right]_{j,k=1,\ldots,N}\rmd
\theta_1\cdots \rmd \theta_N.
\end{split}
\end{equation}
The quantity $Q_N^{(\sigma_1,\sigma_2)}(\theta_1,\theta_2)$ is
called the {\it kernel} of the Haar measure and
\begin{equation}
\label{equalint}
\det\left[f(\theta_j)f(\theta_k)Q^N_{(\sigma_1,\sigma_2)}
\left(\theta_j,\theta_k\right)\right]_{j,k=1,\ldots,N} =
\left(\prod_{j=1}^{N}f(\theta_j)f(-\theta_j)\right)
P^N_{(\sigma_1,\sigma_2)}(\theta_1,\ldots,\theta_N),
\end{equation}
where $P^N_{(\sigma_1,\sigma_2)}(\theta_1,\ldots,\theta_N)$ is the
Haar measure~\eqref{HmeasW}. The integral~\eqref{altfo2} can
always be expressed in terms of the independent eigenvalues;
indeed in appendix~C it is shown that it is always proportional to
the integral
\begin{equation}
\label{avsh}
\begin{split}
\left\langle \prod_{j=1}^N g(\theta) \right \rangle_{\GM}
&=\int_{0}^{\pi}\cdots \int_{0}^{\pi} \left(\prod_{j=1}^N
g(\theta_j)\right)
P^N_{(\sigma_1,\sigma_2)}\left(\theta_j,\ldots,\theta_N\right)\rmd
\theta_1\cdots \rmd \theta_N \\
& =
\det(\alpha^{\left(\sigma_1,\sigma_2\right)}_{jk})_{j,k=0,\ldots,N
-1},
\end{split}
\end{equation}
with a constant of proportionality that depends on the group and
on the function $f(\theta)$. In equation~\eqref{avsh} we have set
$g(\theta)=f(\theta)f(-\theta)$. Explicit expressions for the
matrix elements $\alpha_{jk}^{(\sigma_1,\sigma_2)}$ and the
relations between the averages~\eqref{altfo2} and~\eqref{avsh} for
the various compact groups are reported in table~1, appendix C. In
the rest of the paper we shall concern ourselves only with
integrals of the form~\eqref{avsh}.

Let us now go back to $\ONep$.  In appendix~C we show that
\begin{subequations}
\label{inttras2}
\begin{align}
\alpha_{00} &= g_0\\
\alpha_{0j}& =\alpha_{j0} =  \sqrt{2}g_j, \quad j> 0, \\
\alpha_{jk} & = g_{j-k} + g_{j + k}, \quad j,k
> 0,
\end{align}
\end{subequations}
where for simplicity we have dropped the superscript
$(\sigma_1,\sigma_2)$ and
\begin{equation}
\label{fcoeff} g_l=\frac{1}{2\pi}\int_0^{2\pi}g(\theta)
\rme^{-\rmi l \theta} \rmd \theta
\end{equation}
is the $l$th Fourier coefficient of $g(\theta)$.  Matrices of the
form $\{h_{j+k}\}_{j,k=0,\ldots,N-1}$ are called Hankel matrices,
therefore the matrix $\{\alpha_{jk}\}_{j,k=0,\ldots,N-1}$ is
always the sum of a Toeplitz and a Hankel matrix.

How can we infer from~\eqref{equalint} the structure of the
matrices $\bA$ and $\bB$ that appear in the
Hamiltonian~\eqref{impH}? After all, the geometry of  $H_{\alpha}$
is that of a discrete lattice while the kernel of $\ONep$ lives on
the circle, its explicit form being
\begin{equation}
\label{kernelort2n}
Q^N_{\ONep}(\phi,\psi) = S_{2N-1}(\phi - \psi)
+ S_{2N - 1}(\phi + \psi), \quad \phi,\psi \in [0,\pi),
\end{equation}
where $S_N(z)$ is the kernel~\eqref{UNkernel}. In appendix~C it is
shown that the matrix elements $\alpha_{jk}$ can be expressed as
integral transforms involving a particular class of orthogonal
polynomials, known as Jacobi polynomials. Furthermore, the kernel of Haar
measure can be expressed in the form (see, e.g.,~\cite{Seg59},
p. 24)
\begin{equation}
\label{kort3ndef} Q^N_{\ONep}(\phi,\psi) =
\sum_{j=0}^{N-1}p_j(\cos\phi)p_j(\cos\psi),
\end{equation}
where $p_j(x)$ is the $j$th Chebyshev polynomial of the first
kind:
\begin{equation}
\label{Chpolyft0} p_0(x) = \frac{1}{\sqrt{\pi}} \quad {\rm and }
\quad p_j(x)=\sqrt{\frac{2}{\pi}}\cos\left(j\cos^{-1}x\right),
\quad j
>0.
\end{equation}
Formula~\eqref{kort3ndef} leads to the following expression for
the matrix elements that appear in the determinant of the
left-hand side of~\eqref{equalint}:
\begin{equation}
\label{kort2ndef} f(\phi)f(\psi)Q^N_{\ONep}(\phi,\psi) =
f(\phi)f(\psi)\sum_{j=0}^{N-1}p_j(\cos\phi)p_j(\cos\psi).
\end{equation}
Then, if we compare the integral transforms~\eqref{inttro2n} with
the sum~\eqref{kort2ndef}, we note that the expressions are the
same, but the role of the continuous and discrete variables is
exchanged.  In other words, the functional form of the
kernel~\eqref{kernelort2n} is complementary to the intrinsic
structure of the matrix $\{\alpha_{jk}\}_{j,k=0,\ldots,N-1}$ as a
Toeplitz plus Hankel matrix. It is therefore natural to assume
that the matrices $A$ and $B$ defining the quadratic
form~\eqref{impH} should be the sum of Toeplitz plus Hankel
matrices. As for $\UN$, the periodic boundary conditions will
impose on them a further structure which will turn out to be
essential to our study. It is worth noting that the analysis
presented in section~\ref{unitary} in terms of invariance under
translations of the Haar measure of $\UN$ and of $H_{\alpha}$ is
equivalent to the one discussed here; in the case of $\UN$ the
orthogonal polynomials $p_j(\cos \phi)$ are replaced by the
complex exponentials $\rme^{\rmi j \phi}/\sqrt{2\pi}$.

The above considerations lead one to consider matrices $\bA$ of
the form
\begin{equation}
\label{choice}
\bA_{jl} = a(j-l) + a(j+l), \quad j,l=0,\ldots,M-1.
\end{equation}
Because of the periodic boundary conditions $a$ must be a function
on $\mathbb{Z}/M\mathbb{Z}$, which must also be even in order for
$\bA$ to be symmetric. Clearly, a Hankel matrix cannot be
antisymmetric, therefore $\gamma$ must be zero: the
Hamiltonian~\eqref{impH} must be isotropic.  A brief look to
table~1 in appendix~C shows that this is a necessary condition for
all the other compact groups too.  Since $\bA$ is a real symmetric
matrix, its eigenvalues $\Lambda_k$ are real and therefore
\begin{equation}
\label{psiort}
\vpsi_k = \frac{\Lambda_k}{\abs{\Lambda_k}}\vphi_k
= \sign \Lambda_k \vphi_k,
\end{equation}
where the $\vphi_k$s are the eigenvectors of $\bA$.

We now need to diagonalize $\bA$; as for the unitary group, this
can be done explicitly.  Because
\begin{equation}
\label{symor}
\bA_{jl}=\bA_{j\, M -l},
\end{equation}
 any odd function on $\mathbb{Z}/M\mathbb{Z}$ will
 be in the kernel of $\bA$. Therefore,
\begin{equation}
\label{seig} \phi_{kj} = \sqrt{\frac{2}{M}}\sin k j, \quad
k=\frac{2\pi l}{M}, \quad l =1,\ldots,[(M-1)/2],
\end{equation}
are a set of independent eigenvectors with eigenvalue zero whose
 multiplicity is at least $[(M-1)/2]$, where $[\cdot]$ denotes the
integer part.

The eigenvectors with non-zero eigenvalues can be found by
exploiting the symmetries of the matrix~\eqref{choice}. For
simplicity, we assume that the non-zero eigenvalues are
non-degenerate. An immediate consequence of~\eqref{symor} and of
the condition $\Lambda_k\neq 0$ is that any eigenvector $c_k(j)$
must be an even function on $\mathbb{Z}/M\mathbb{Z}$; thus, we can
always write
\begin{equation}
\label{even}
c_k(j) = \rme_k(j) + \rme_k(-j), \quad j \in
\mathbb{Z}/M\mathbb{Z}.
\end{equation}
 Furthermore, using the periodicity of $a(j)$, it is easy to show that
 if $c(j)$ is an eigenvector of $\bA$, than $c_k(j+p) + c_k(j - p)$, where $p$ is
an arbitrary integer, is also an eigenvector corresponding to the
same eigenvalue $\Lambda_k$. It follows that
\begin{equation}
\label{mult1}
c_k(j+p) + c_k(j - p) \propto \rme'_k(p)c_k(j).
\end{equation}
Since $c_k(j)$ is even, the role of $j$ and $p$ can be
interchanged, and since they are both arbitrary, we can choose
$\rme'_k(j)=\rme_k(j)$.  Then it follows from
equations~\eqref{even} and~\eqref{mult1} that
\begin{equation}
\label{addmult2}
\rme_k(j+p) = \rme_k(j)\rme_k(p),
\end{equation}
for an appropriate choice of the constant of proportionality
in~\eqref{mult1}.  Thus, the $\rme_k(j)$ are
additive-multiplicative functions.  Because they are periodic too,
it must be that
\begin{equation}
\rme_k(M) = \rme_k(1)^M = 1.
\end{equation}
Therefore the $\rme_k(j)$s are roots of unity:
\begin{equation}
\rme_k(j) = \frac{ \rme^{\rmi kj}}{\sqrt{M}},\quad k=\frac{2\pi
l}{M}, \quad j,l=0,\ldots M-1.
\end{equation}
As immediate consequence of~\eqref{even} the remaining normalized
eigenvectors of matrix~\eqref{choice} are
\begin{equation}
\label{coseig}
\begin{cases}
\phi_{0j} = \frac{1}{\sqrt{M}},\quad \phi_{kj} =
\sqrt{\frac{2}{M}}\cos k j, \quad 0<k <\pi & \text{for $M$ odd}\\
\phi_{0j} = \frac{1}{\sqrt{M}},\quad \phi_{kj} =
\sqrt{\frac{2}{M}}\cos k j, \quad 0<k <\pi,\quad \phi_{\pi
j}=\frac{(-1)^j}{\sqrt{M}} & \text{for $M$ even,}
\end{cases}
\end{equation}
where $k = 2\pi l/M$. The corresponding eigenvalues can be
obtained by direct substitution:
\begin{equation}
\label{fexp}
\Lambda_k = \begin{cases} 2a(0) + 4\sum_{j =
1}^{(M-1)/2}a(j)\cos
kj & \text{if $M$ is odd} \\
 2\left[a(0) + (-1)^l a(M/2)\right] + 4\sum_{j = 1}^{M/2-1}a(j)\cos
kj & \text{if $M$ is even.}
\end{cases}
\end{equation}

In appendix~B we show that there exists a canonical transformation
of the Fermi operators $b_j$ that diagonalizes~\eqref{impH}. Using
the same notation as in appendix~B, let us denote by $\eta_k$ the
Fermi operators in term of which $H_{\alpha}$ is diagonal. The
fact that approximately half of the $\Lambda_k$s are zero means
that the corresponding $\eta_k$s do not appear in $H_{\alpha}$. In
other words, $H_{\alpha}$ is isomorphic to a system with half the
number of degrees of freedom.  This is not surprising; it is a
reflection of the fact that only half of the eigenvalues of a
matrix in $\ONep$ are independent.  In the same way as statistical
properties of orthogonal and symplectic matrices are computed only
in terms of the independent eigenvalues, so the extra degrees of
freedom in $H_{\alpha}$ can be ignored.  The matrix~\eqref{Tmatr}
can therefore be determined from the eigenvectors~\eqref{coseig}.

Following the same steps as for $\UN$ we fix our attention on the
subsystem P composed of the first $N$ consecutive oscillators and
let $M \rightarrow \infty$. The eigenvalues~\eqref{fexp} converge
to the even function
\begin{equation}
\label{lambdathet}
\Lambda(\theta) = \Lambda_0 + 2
\sum_{j=1}^{\infty}\Lambda_j \cos\theta j,
\end{equation}
where $\Lambda_j=2a(j)$.  Finally, by substituting the
vectors~\eqref{coseig} into~\eqref{Tmatr} and taking the limit
$M\rightarrow \infty$, we obtain
\begin{equation}
\label{vgc} (T_N)_{jl}=  \frac{2}{\pi}\sum_{k=0}^{\pi}
\frac{\Lambda_k}{\abs{\Lambda_k}}
 \cos{k j}\cos{k l}\Delta k \xrightarrow[M \rightarrow \infty]{}
 \alpha_{jk},
\end{equation}
where the $\alpha_{jk}$ are precisely those of
equation~\eqref{inttras2} with symbol
\begin{equation}
g(\theta)= \frac{\Lambda(\theta)}{\abs{\Lambda(\theta)}}.
\end{equation}
We can then define an XX model with orthogonal symmetry by
choosing
\begin{equation}
a(j) = \alpha\left[\delta_1(j) + \delta_1(-j)\right] -2
\delta_0(j).
\end{equation}

It is important to notice that in order for $T_N[g]$ to be the sum
of a Toeplitz and a Hankel matrix the subchain P must be made of
the first $N$ sequential oscillators: we cannot shift the
subsystem $P$ because  $H_{\alpha}$ is not translation invariant.
This property was to be expected because the
kernel~\eqref{kernelort2n} is not invariant under translations. In
other words, the origin of the lattice defining the spin chain is
a privileged point, in the same way as the point $1$ on the unit
circle is a symmetry point for the spectra of orthogonal and
symplectic matrices.  A consequence of the absence of
translational invariance is that the determinant expressing the
spin-spin correlations~\eqref{Wick-Th} is an average over $\ONep$
only if one of the spins is the first in the chain.

It turns out that the structure of the matrix $T_N[g]$ so obtained
has important consequences for the formula~\eqref{JKidea2}. First,
since $\Lambda(-\theta)=\Lambda(\theta)$, $T_N[g]$ is symmetric,
therefore the correlation matrix $C_N$ factorizes as
in~\eqref{toepf}.  Thus, the characteristic polynomial
\begin{equation}
D_N[g](\lambda) = \det\left(\lambda I - T_N[g]\right)
\end{equation}
 in the integral~\eqref{JKidea2} is
an average over $\ONep$.  We compute this
 integral in section~\ref{Fisher-Hartwig}.  The symbol
 $\Lambda(\theta)/\abs{\Lambda(\theta)}$ is the same as the one
 discussed in section~\ref{unitary} for the case when $\gamma =0$:
 it is a piecewise continuous function that takes the values
$1$ and $-1$ and whose jumps are located at the points $\theta_r$
which are solutions of equation~\eqref{jloc}.  In
section~\ref{unitary} we have seen that the eigenvalues of the
corresponding Toeplitz matrix are in the interval $[-1,1]$;
similar arguments lead to the same conclusion for the eigenvalues
of a matrix which is the sum of a Toeplitz and a Hankel matrix
with the same symbol.  Therefore, formula~\eqref{JKidea2} gives
the entropy of the subchain P and can be expressed in term of an
average over $\ONep$.

\section{$\SptN$ and $\ONen$ symmetry}
\label{symplectic}

The treatment of these two groups turns out to be the same --
see~\eqref{leq}. The arguments are analogous to those presented
for $\ONep$. The elements in $\SptN$ are $2N \times 2N$ unitary
matrices $U$ such that
\begin{equation}
U J U^t = J,\quad J=\begin{pmatrix} 0 & -I \\ I & 0 \end{pmatrix},
\end{equation}
where $I$ is the $N\times N$ identity matrix.  The number of
independent eigenvalues in both $\ONen$ and $\SptN$ is $N$.
Without loss of generality, we shall concentrate only on $\SptN$.
The kernel of the Haar measure
\begin{equation}
\label{ksymp}
 Q^N_{\SptN}(\phi,\psi) = S_{2N+1}(\phi - \psi) -
S_{2N + 1}(\phi + \psi), \quad \phi,\psi \in [0,\pi),
\end{equation}
and the matrix elements appearing in the determinant~\eqref{avsh}
are
\begin{equation}
\label{melSp}
\alpha_{jk}=g_{j-k} - g_{j+k +2}.
\end{equation}

As for $\ONep$ the structure of formulae~\eqref{ksymp}
and~\eqref{melSp} is incompatible with $\gamma \neq 0$. The choice
of the matrix $\bA$ is
\begin{equation}
\label{bAsym}
\bA_{jk} = a(j-k) - a(j+k +2),
\end{equation}
where $a$ is an even function on $\mathbb{Z}/M\mathbb{Z}$. The
diagonalization of the matrix~\eqref{bAsym}  is analogous to the
one of~\eqref{choice}. Therefore, we just present the results. The
eigenvectors that span the kernel of $\bA$ are
\begin{subequations}
\begin{equation}
\label{coseig2}
\phi_{kj} = \sqrt{\frac{2}{M}} \cos k (j + 1),
\quad k=\frac{2\pi l}{M},
\quad l =0,\ldots,[M/2],\\
\end{equation}
while those corresponding to $\Lambda_k \neq 0$ are
\begin{equation}
\label{sineig2}
 \phi_{kj} = \sqrt{\frac{2}{M}}\sin k (j +1),
\quad k=\frac{2\pi l}{M}, \quad l =1,\ldots,[(M-1)/2].
\end{equation}
\end{subequations}
As in the case of $\ONep$ the Hamiltonian~\eqref{impH} is
isomorphic to a system with half the number of degrees of freedom.
The eigenvectors associated to the relevant degrees of freedom are
those in~\eqref{sineig2}.  Similarly, the eigenvalues can be
computed by direct substitution; they turn out to be given by
formula~\eqref{fexp}. By fixing the number $N$ of oscillators in
the subchain P and letting $M \rightarrow \infty$, the matrix
$T_N$ converges to
\begin{equation}
\label{vguuffa}
\begin{split}
(T_N)_{jl}& = \frac{2}{\pi}\sum_{k=0}^{\pi}
\frac{\Lambda_k}{\abs{\Lambda_k}}
 \sin{k (j+1)}\sin{k (l+1)}\Delta k \\
 & \quad   \xrightarrow[M \rightarrow \infty]{}
 \frac{1}{2\pi}\int_0^{2\pi}\frac{\Lambda(\theta)}{\abs{\Lambda(\theta)}}\left(
\rme^{-\rmi \left(j - l\right)\theta} - \rme^{-\rmi\left(j + l +
2\right)\theta} \right)\rmd \theta,
\end{split}
\end{equation}
which are the integral transforms~\eqref{symal}; $\Lambda(\theta)$
is the same real and even function as in~\eqref{lambdathet},
therefore $T_N[g]$ is symmetric.  An immediate consequence is that
the entropy formula~\eqref{JKidea2} can be expressed in terms of
an average over $\SptN$.

\section{$\ONo{\pm}$ symmetry}
\label{ortodd}

The treatments of $\ONo{+}$ and of $\ONo{-}$ follow a similar
pattern, indeed averages over these two groups are intertwined by
equations~\eqref{ONeav} and~\eqref{ONeavuf}. The kernels of the
Haar measures are
\begin{equation}
\label{kono}
 Q^N_{\ONo{\pm}}(\phi,\psi) = S_{2N}(\phi - \psi) \mp
S_{2N}(\phi + \psi), \quad \phi,\psi \in [0,\pi),
\end{equation}
and the matrix elements in the average~\eqref{avsh} are
\begin{equation}
\label{onmav}
\alpha_{jk} = g_{j-k} \mp g_{j+k+1},
\end{equation}
where in equations~\eqref{kono} and~\eqref{onmav} the minus sign
refers to $\ONo{+}$ and the plus sign to $\ONo{-}$. Consequently,
the choices of the matrix $\bA$ compatible with~\eqref{kono}
and~\eqref{onmav} are
\begin{equation}
\label{acorto}
\bA_{jk} = a(j-k) \mp a(j+k+1),
\end{equation}
where $a$ is an even function on $\mathbb{Z}/M\mathbb{Z}$.  As
with the groups treated previously, equations~\eqref{kono}
and~\eqref{onmav} are incompatible with $\gamma \neq 0$.

The matrix~\eqref{acorto} can be diagonalized adopting the same
techniques used for the other groups.
\begin{subequations}
\label{seteig3}
\begin{align}
\label{sineig3}
 \phi_{kj} &= \sqrt{\frac{2}{M}}
 \sin\left[ k \left(\frac{2j +1}{2}\right)\right],
\quad k=\frac{2\pi l}{M}, \quad l =1,\ldots,[M/2], \\
\label{coseig3} \phi_{kj}&= \sqrt{\frac{2}{M}} \cos \left[k
\left(\frac{2j + 1}{2}\right)\right], \quad k=\frac{2\pi l}{M},
 \quad l =0,\ldots,[(M-1)/2].
\end{align}
\end{subequations}
 These are the eigenvectors of the matrix~\eqref{acorto} for
both choice of signs; however, the functions~\eqref{coseig3} are
in the kernel of $\bA$ when the sign between the two terms
in~\eqref{acorto} is minus, i.e. for $\ONo{+}$ symmetry, while
their eigenvalues are not zero when the sign is plus, i.e. for
$\ONo{-}$ symmetry. For the eigenvectors~\eqref{sineig3} the role
is reversed.  The eigenvalues are given by formula~\eqref{fexp}
for these groups too.  It is now straightforward to determine the
matrices $T_N$ for both groups:
\begin{subequations}
\begin{equation}
\begin{split}
(T_N)_{jl}& = \frac{2}{\pi}\sum_{k=0}^{\pi}
\frac{\Lambda_k}{\abs{\Lambda_k}}
 \sin{\left[k \left(\frac{2j+1}{2}\right)\right]}
 \sin{\left[k \left(\frac{2l+1}{2}\right)\right]}
 \Delta k \\
& \quad  \xrightarrow[M \rightarrow \infty]{}
 \frac{1}{2\pi}\int_0^{2\pi}\frac{\Lambda(\theta)}{\abs{\Lambda(\theta)}}\left(
\rme^{-\rmi \left(j - l\right)\theta} - \rme^{-\rmi\left(j + l +
1\right)\theta} \right)\rmd \theta
\end{split}
\end{equation}
for $\ONo{+}$ and
\begin{equation}
\begin{split}
(T_N)_{jl}& =  \frac{2}{\pi}\sum_{k=0}^{\pi}
\frac{\Lambda_k}{\abs{\Lambda_k}}
 \cos{\left[k \left(\frac{2j+1}{2}\right)\right]}\cos{\left[k \left(\frac{2l+1}{2}\right)\right]}
 \Delta k \\
& \quad  \xrightarrow[M \rightarrow \infty]{}
 \frac{1}{2\pi}\int_0^{2\pi}\frac{\Lambda(\theta)}{\abs{\Lambda(\theta)}}\left(
\rme^{-\rmi \left(j - l\right)\theta} + \rme^{-\rmi\left(j + l +
1\right)\theta} \right)\rmd \theta
\end{split}
\end{equation}
\end{subequations}
for $\ONo{-}$. The function $\Lambda(\theta)$ is even and the
symbol $g(\theta) = \Lambda(\theta)/\abs{\Lambda(\theta)}$ is the
same function analyzed in connection with the other groups.  The
matrix $T_N[g]$ is real and symmetric and therefore the
formula~\eqref{JKidea2} for the entropy of the entanglement has an
interpretation as average over $\ONo{+}$ or over $\ONo{-}$.

\section{Generalizations of the Fisher-Hartwig formula and the
computation of entanglement}
\label{Fisher-Hartwig}

The computation of Toeplitz determinants, and in particular of
their asymptotics, is important in many branches of Physics. The
first and most famous application goes back to 1946 and is due to
Osanger, who showed that the diagonal spin-spin correlations in
the classical two-dimensional Ising model can be expressed in
terms of Toeplitz determinants.  It turns out that the behaviour
of the leading order term as the dimension of the matrix tends to
infinity changes radically when the symbol has discontinuities or
zeros.  Indeed, phase transitions in quantum and classical lattice
systems often appear as changes in the analytic properties of
symbols in Toeplitz determinants.

When the symbol $g(\theta)$ is a continuous function on the unit
circle and the $c_k$s are the Fourier coefficients of $\log
g(\theta)$, Szeg\H{o}'s theorem~\cite{Sze52} states that
\begin{equation}
\label{Szego} \ln \det T_N[g] = c_0N + \sum_{k=1}^\infty
kc_kc_{-k} + {\rm o}(1), \quad N \rightarrow \infty,
\end{equation}
provided that the series $\sum_{k=-\infty}^{\infty}\abs{c_k}$ and
$\sum_{k=-\infty}^{\infty}\abs{k}\abs{c_k}^2$ converge.  If
$g(\theta)$ has zeros or discontinuities, then it can always be
reduced to the form
\begin{equation}
\label{FHsymb}
g(\theta) = \phi(\theta)\prod_{r=1}^L
u_{\alpha_r}(\theta - \theta_r)t_{\beta_r}(\theta-\theta_r),
\end{equation}
where $\phi$ is smooth, has winding number zero and
\begin{subequations}
\begin{align}
t_\beta(\theta) & = \exp\left[ - \rmi\beta\left(\pi -
\theta\right)\right], \quad 0 \le \theta < 2\pi, \quad \beta \not \in \mathbb{Z}\\
u_\alpha(\theta) & = (2 - 2\cos \theta)^\alpha,\quad \rpart \alpha
> -\frac12.
\end{align}
\end{subequations}
Note that $L$ represents the number of zeros/discontinuities in
the interval $[0, 2\pi)$. Fisher and Hartwig~\cite{FH68}
conjectured that
\begin{equation}
\label{FHconj} \ln D_N[g] =   c_0N + \left(\sum_{r=1}^L \alpha_r^2
- \beta_r^2\right)\ln N  + \ln E + {\rm o}(1), \quad N \rightarrow
\infty,
\end{equation}
where now the $c_k$s are the Fourier coefficients of $\ln
\phi(\theta)$.  Basor~\cite{Bas78} determined the constant $E$:
\begin{equation}
\label{costE}
\begin{split}
E & = \exp\left( \sum_{k=1}^\infty kc_kc_{-k}\right) \prod_{r=1}^L
\phi_+\left(\rme^{\rmi \theta_r}\right)^{-\left(\alpha_r +
\beta_r\right)}\phi_{-}\left(\rme^{-\rmi
\theta_r}\right)^{-\left(\alpha_r - \beta_r\right)}\\
& \quad \times \prod_{1\le r \neq s\le L}\left(1 -
\exp\left[\rmi\left(\theta_s -
\theta_r\right)\right]\right)^{-\left(\alpha_r +
\beta_r\right)\left(\alpha_s -
\beta_s\right)}\prod_{r=1}^L\frac{\G\left(1 + \alpha_r +
\beta_r\right)\G\left(1 + \alpha_r - \beta_r\right)}{\G\left(1 +
2\alpha_r\right)},
\end{split}
\end{equation}
where $\G(z)$ is the {\it Barnes G-function}\footnote{The
definition of the Barnes G-function is
\[
\G(z) = \left(2\pi\right)^{z/2}\rme^{-\left[z(z+1) + \gamma_{{\rm E}}
z^2\right]/2}\prod_{n=1}^\infty \left[\left(1 +
\frac{z}{n}\right)^n\rme^{-z + z^2/(2n)}\right],
\]
where $\gamma_{{\rm E}}$ is Euler's constant.  It generalizes the Gamma
function, in the sense that it obeys $\G(z+1)=\Gamma(z)\G(z)$.}
and
\begin{equation}
\ln \phi_+(t) = \sum_{j=1}^{\infty}c_jt^j, \quad \ln \phi_-(t) =
\sum_{j=1}^\infty c_{-j}t^{-j}.
\end{equation}
The Fisher-Hartwig conjecture has been proved for $\abs{\rpart
\alpha_r} < 1/2$ and $\abs{\rpart \beta_r}< 1/2$~\cite{BS86} and
for other specific values of $\alpha_r$, $\beta_r$ and $L$. (The
reader is referred to~\cite{BS90}, pp. 469--474, for a complete
discussion.)  The most important difference between
equations~\eqref{Szego} and~\eqref{FHconj} is the extra term
proportional to $\ln N$ in the Fisher-Hartwig formula.

Formula~\eqref{FHconj} was used by Jin and Korepin~\cite{JK03} to
compute the entropy of the entanglement for the XX model, that is
for the Hamiltonian~\eqref{XYmodel} with $\gamma=0$.  Their
computation can be easily generalized to all Hamiltonians of the
form~\eqref{impH} that are invariant under translations and
isotropic,  for which, therefore, the formula~\eqref{JKidea2}
becomes an average over $\UN$. However, before entering in the
details of the computation, we discuss the generalization of
formula~\eqref{FHconj} to determinants of combinations of Toeplitz
and Hankel matrices, or, more specifically, determinants that can
be interpreted as averages over the other classical compact
groups.

Basor and Ehrhardt~\cite{BE02} proved a generalization of the
Fisher-Hartwig formula to determinants of matrices of the type
\begin{equation}
g_{j-k} + g_{j+k+1}, \quad j,k =0, \ldots, N-1,
\end{equation}
i.e. determinants that are averages over $\ONo{-}$.  Using
equations~\eqref{ONeav} and~\eqref{ONeavuf}, their formulae can be
applied to averages over $\ONo{+}$ by a change of variable in the
integral~\eqref{equalint}.  In a recent paper Forrester and
Frankel~\cite{FF04} extended the results of Basor and Ehrhardt
conjecturally to averages over $\SptN$, $\ONep$ and $\ONen$. These
formulae provide a very efficient means to compute the
integral~\eqref{JKidea2} for chains of fermionic oscillators whose
symmetries are associated to one of these groups.

From the discussion of the previous sections it has emerged that
we are only interested in the symbol
$g(\theta)=\Lambda(\theta)/\abs{\Lambda(\theta)}$ when
$\Lambda(\theta)$ is real. This symbol takes only two values: $1$
and $-1$. Its discontinuities are located at the points where
equation~\eqref{jloc} has solutions. If such an equation has no
solutions, then $g(\theta)$ is a constant and  $T_N[g] = \pm I$
for all the compact groups.  It follows from equation~\eqref{nent}
that $E_{\rm P}=0$. In physical terms this means that the
Hamiltonian $H_{\alpha}$ is away from the critical point or,
equivalently, the magnetic field is so strong that all the spins
are aligned, thus there are no correlations and correspondingly
the entanglement must be zero.

The Fisher-Hartwig formula extended to all the compact groups is
fairly complicated when expressed for a general symbol of the
form~\eqref{FHsymb}. For simplicity, we shall report only the
expression for the case that concerns us directly, i.e.
$g(\theta)$ is even and has only discontinuities; for the general
case we refer to~\cite{BE02}.  Equation~\eqref{FHsymb} becomes
\begin{equation}
\label{evensymbol} g(\theta)=\phi(\theta)\prod_{r=1}^R
t_{\beta_r}(\theta - \theta_r) t_{-\beta_r}(\theta + \theta_r),
\end{equation}
where now all the discontinuities $\theta_r$ lie in the interval
$[0, \pi)$ and therefore $L=2R$; we exclude the case $\theta_r =
0,\pi$ and still require $\abs{\rpart \beta_r}<1/2$.  We have
\begin{equation}
\label{FHext}
\begin{split}
\ln D_N[g](\lambda)_{\lamp} & = Nc_0 - \left(\sum_{r=1}^R
\beta_r^2\right)\ln N  \\
& \quad + \frac12\sum_{k=1}^\infty k c_k^2 + \sum_{k=1}^\infty
c_{2k-1} +  \ln F_{\lamp} + \ln E + {\rm o}(1), \quad N
\rightarrow \infty,
\end{split}
\end{equation}
where
\begin{subequations}
\label{cgterms}
\begin{align}
\label{gdepterm}
F_{(\sigma_1,\sigma_2)} &= \prod_{r=1}^R(1 + \rme^{\rmi
\theta_r})^{(\sigma_1 -1/2)\beta_r}(1 + \rme^{-\rmi \theta_r
})^{-(\sigma_1 -1/2)\beta_r} (1 + \rme^{\rmi
\theta_r})^{\left(\sigma_2 + 1/2\right)\beta_r}(1 + \rme^{-\rmi
\theta_r})^{-\left(\sigma_2 + 1/2\right)\beta_r} \nonumber\\
& \quad \times \exp\left\{-\sum_{k=1}^\infty c_k
\left[\sigma_1 -1/2 + (-1)^k\left(\sigma_2 + 1/2\right)\right]\right\}\\
\label{cterm}
 E  & = \prod_{r=1}^R 2^{-\beta_r^2}\G(1 +
\beta_r)\G(1-\beta_r)\abs{1 - \rme^{2\rmi \theta_r}}^{-\beta_r^2}
\frac{(1 - \rme^{-\rmi \theta_r})^{\beta_r/2}(1-  \rme^{\rmi
\theta_r})^{-\beta_r/2}}{(1+
\rme^{-\rmi \theta_r})^{\beta_r/2}(1+\rme^{\rmi \theta_r})^{-\beta_r/2}} \nonumber \\
& \quad \prod_{1 \le r < s \le R} \left |\frac{1 -
\rme^{\rmi\left(\theta_r - \theta_s\right)}}{1 - \rme^{\rmi
\left(\theta_r + \theta_s\right)}}\right|^{2\beta_r\beta_s}
\prod_{r=1}^R\phi_+\left(\rme^{\rmi
\theta_r}\right)^{\beta_r}\phi_-\left(\rme^{\rmi
\theta_r}\right)^{-\beta_r}.
\end{align}
\end{subequations}

Note that the main differences between the various groups appear
only in the ${\rm O}(1)$ terms of~\eqref{FHext}: when an even
symbol with discontinuities is averaged over a compact group the
term linear in $N$ in formulae~\eqref{FHconj} and~\eqref{FHext} is
the same in every case, and the term logarithmic in $N$ has just
an extra factor of two in front of it for $\UN$.  This factor is
due to the fact that averages over $\UN$ are computed by
integrating over $[0,2\pi)^N$, while for the other compact groups
the domain of integration is $[0,\pi)^N$, therefore, in the latter
case the singularities located at $-\theta_r$ do not contribute.

The representation~\eqref{evensymbol} of
$g(\theta)=\Lambda(\theta)/\abs{\Lambda(\theta)}$ is given by the
following choices of $\phi(\theta)$ and $\beta_r$:
\begin{subequations}
\begin{align}
\label{chc}
\phi(\lambda) & = (\lambda + 1)\left(\frac{\lambda +
1}{\lambda -1}\right)^{\left(\sum_{r=1}^R
(-1)^r\theta_r\right)/\pi},\\
 \beta_r(\lambda)& = (-1)^r \beta(\lambda), \quad
 \beta(\lambda)  = \frac{1}{2\pi
\rmi}\ln \left(\frac{\lambda + 1}{\lambda - 1}\right),
\end{align}
\end{subequations}
with $-\pi \le \arg[ (\lambda+1)/(\lambda-1) ]< \pi $.  Therefore,
we have $\abs{\rpart \beta_r} < 1/2$ on the contour of integration
$c(\epsilon,\delta)$ and we can apply
formula~\eqref{FHext}. The leading order asymptotic of the
entropy~\eqref{JKidea2} is then given by
\begin{equation}
E_{\rm P} = I_1\, N  - 2^{w_\G} R\, I_2 \, \ln N + {\rm O}(1), \quad N
\rightarrow \infty,
\end{equation}
where $R$ is the number of discontinuities in the interval $[0, \pi)$
and
\begin{equation}
w_{\G} = \begin{cases} 1 & \text{if the average is over $\UN$} \\
                        0 & \text{if the average is over the other
                                   compact groups.}
          \end{cases}
\end{equation}
$I_1$ and $I_2$ are the integrals
\begin{subequations}
\begin{align}
I_1 & = \lim_{\epsilon \rightarrow 0^+} \lim_{\delta \rightarrow
0^+} \frac{1}{2\pi \rmi} \oint_{c(\epsilon,\delta)} \rme(1 +
\epsilon,\lambda) \frac{\phi'(\lambda)}{\phi(\lambda)} \rmd
\lambda =0, \\
\label{JK1}
I_2 & = \lim_{\epsilon \rightarrow 0^+} \lim_{\delta \rightarrow
0^+} \frac{1}{\pi \rmi} \oint_{c(\epsilon,\delta)} \rme(1 +
\epsilon,\lambda) \beta(\lambda)\beta'(\lambda) \rmd \lambda = -1/(6\ln 2).
\end{align}
\end{subequations}
The first integral can be computed straightforwardly using the
residue theorem; the second was computed in~\cite{JK03} (in this
case the integrand is a multivalued function inside the contour).
Finally, we have
\begin{equation}
\label{mainresult}
 E_{\rm P} \sim \frac{2^{w_{\G}} R}{6}\log_2 N, \quad N
\rightarrow \infty.
\end{equation}

The asymptotic relation~\eqref{mainresult} represents one of our
main results. As mentioned in section~\ref{XYpres}, the
logarithmic growth of the entanglement is a general consequence of
the fact that in one dimensional systems near quantum phase
transitions the entropy is a logarithmic function of the size of
the system~\cite{Kor04,CC04}. These thermodynamic arguments give
the value of the coefficient in front of the logarithm to be
one-third of the central charge of the associated Virasoro
algebra.  Our expression therefore leads to an explicit formula
for the central charge, which depends in a non-trivial way on the
geometry of the Hamiltonian. The factor $2^{w_{{\rm G}}}$ is
universal, depending only on the symmetries determining the
classical compact group to be averaged over.  The factor $R$
corresponds to the number of solutions of~\eqref{jloc}.

We now proceed to compute the next-to-leading-order term.  This
can be determined by integrating with respect to $\lambda$ the
terms independent of $N$ in the Fisher-Hartwig
formulae~\eqref{FHconj} and~\eqref{FHext}.  We begin with $\UN$.

Since the symbol is even, using~\eqref{evensymbol}
and~\eqref{chc}, the constant~\eqref{costE} becomes
\begin{equation}
\label{newevenE}
\begin{split}
E(\lambda) & = \prod_{r=1}^R\abs{1 - \rme^{\rmi
2\theta_r}}^{-2\beta(\lambda)^2} \prod_{1 \le r < s \le R} \left
|\frac{1 - \rme^{\rmi\left(\theta_r - \theta_s\right)}}{1 -
\rme^{\rmi \left(\theta_r + \theta_s\right)}}\right|^{4(-1)^{(r +
s)}\beta(\lambda)^2} \\
& \quad \times  \left[\G\left(1 + \beta(\lambda)\right)\G\left(1 -
\beta(\lambda)\right)\right]^{2R}
\end{split}
\end{equation}
where only the independent discontinuities, located in the
interval $[0,\pi)$, are taken into account. The logarithmic
derivative of~\eqref{newevenE} is the sum of two terms. The first
one is
\begin{equation}
\label{ftermo} 4 \beta(\lambda)\beta'(\lambda)\left(2\sum_{1 \le r
< s \le R} (-1)^{(r + s)} \ln \left |\frac{1 -
\rme^{\rmi\left(\theta_r - \theta_s\right)}}{1 - \rme^{\rmi
\left(\theta_r + \theta_s\right)}}\right| - \sum_{r=1}^R\ln\abs{1
- \rme^{\rmi 2 \theta_r}} \right)
\end{equation}
The second term is more delicate; we have
\begin{equation}
\label{logd}
\G\left(1 +
\beta(\lambda)\right)\G\left(1 -
\beta(\lambda)\right) = \rme^{-\beta(\lambda)^2(1 + \gamma_{{\rm E}})}\prod_{n=1}^{\infty}
\left(1 - \frac{\beta(\lambda)^2}{n^2}\right)^n\rme^{\beta(\lambda)^2/n},
\end{equation}
where $\gamma_{{\rm E}}$ is Euler's constant.  The logarithmic
derivative of the right-hand side of~\eqref{logd} is
\begin{equation}
\label{stermo}
-2\beta(\lambda)\beta'(\lambda)\left[1 + \gamma_{{\rm E}} + \Upsilon(\lambda)\right],
\end{equation}
where
\begin{equation}
\Upsilon(\lambda)= \sum_{n=1}^\infty \frac{\beta(\lambda)^2/n}%
{n^2 - \beta(\lambda)^2}.
\end{equation}
Let us define
\begin{equation}
K = 1 + \gamma_{{\rm E}} + \frac{1}{R}\left(\sum_{r=1}^R\ln\abs{1
- \rme^{\rmi 2 \theta_r}} -2\sum_{1 \le r < s \le R} (-1)^{(r +
s)} \ln \left |\frac{1 - \rme^{\rmi\left(\theta_r -
\theta_s\right)}}{1 - \rme^{\rmi \left(\theta_r +
\theta_s\right)}}\right| \right)
\end{equation}
and
\begin{equation}
I_3 = \lim_{\epsilon \rightarrow 0^+} \lim_{\delta \rightarrow
0^+} \frac{1}{\pi \rmi} \oint_{c(\epsilon,\delta)} \rme(1 +
\epsilon,\lambda) \beta(\lambda)\beta'(\lambda) \Upsilon (\lambda) \rmd \lambda.
\end{equation}
This integral was evaluated in~\cite{JK03} where it was found that
$I_3=0.0221603...$.

Combining equations~\eqref{newevenE},~\eqref{ftermo}
and~\eqref{stermo}, we obtain that the ${\rm O}(1)$ contribution
to the entropy of entanglement is
\begin{equation}
\label{O1term} C_{\UN} = 2R I_2 K -  2RI_3 = \frac{R}{3\ln2}\left(
K - 6 I_3 \ln2\right).
\end{equation}
When $R=1$ this equation reduces to the result of Jin and Korepin
for the XX model.

Let us now consider the other compact groups.  From
equations~\eqref{cgterms} it is clear that the
next-to-leading-order term is composed of two parts, one, common
to all groups, coming from equation~\eqref{cterm} and the other,
depending on the particular choice of the group, given
by~\eqref{gdepterm}.  By taking the logarithmic derivative of
equation~\eqref{cterm} one immediately realizes that only two
integrals contribute to this term: the first one is $I_2$; the
second one is
\begin{equation}
\label{stint}
I_4 = \frac{1}{2\pi \rmi} \oint_{c(\epsilon,\delta)}
\rme(1 + \epsilon,\lambda) \beta'(\lambda) \rmd \lambda.
\end{equation}
This integral can be evaluated using the residue theorem to obtain
$I_4=0$.  Therefore, by proceeding as for $\UN$ we can determine
the contribution to the sub-leading term coming from
equation~\eqref{cterm}:
\begin{equation}
\label{ntlotg} C_{{\rm G}} = \frac{R}{6\ln2}\left( K - 6I_3\ln 2 +
\ln2 \right).
\end{equation}
This expression differs from that for $C_{\UN}$ by a factor $1/2$,
the origin of which is the same as in the leading order term, and
an additional term equal to $R/6$.

We are left to determine the contribution coming from
equation~\eqref{gdepterm}. It is evident that, up to a constant
depending only on the $\theta_r$s, by taking the logarithmic
derivative of the right-hand-side we are left just with integrals
of the type~\eqref{stint}. Hence, the next-to-leading-order term
for the groups $\SpN$ and $\mathrm{O}^\pm(N)$ is
simply~\eqref{ntlotg}.

\section{Painlev\'e VI and gap probability generating
functions}
\label{Painleve}

The averages we have discussed turn out to be related to solutions
of integrable second order ODEs of Painlev\'e type. There exist
six Painlev\'e equations; any second order differential equation
free of moveable essential singularities of the form
\begin{equation}
y''=R(y',y,t),
\end{equation}
where $R$ is a rational function, can be reduced to one of them.

Let us consider a generating function
$g(\theta)=\Lambda(\theta)/\abs{\Lambda(\theta)}$ with only two
discontinuities at $\theta_1$ and $-\theta_1$ and set $\phi=
2\theta_1$.  We then look at the characteristic
polynomial~\eqref{chpoln} when $g(\theta)$ is averaged over $\UN$,
which can be written as the integral
\begin{equation}
\label{fdef}
\begin{split}
 D_N[g](\lambda)& = \frac{(\lambda + 1)^N}{(2\pi)^N N!}\int_0^{2\pi} \cdots
\int_0^{2\pi}\prod_{j=1}^N \left[1 - \xi \chi_{I_{[\pi - \phi,
\pi)}}(\theta_j)\right]\, \prod_{1\leq j < k \leq N}\left|\,
\rme^{\rmi \theta_j} - \rme^{\rmi \theta_k}\right|^2 \rmd
\theta_1\cdots \rmd \theta_N  \\
& =(\lambda + 1)^N\, \EgU(I_{[\pi-\phi,\pi)};\xi),
\end{split}
\end{equation}
where we have set $\xi= 2/(\lambda + 1)$. The function
$\EgU(I_{[\pi -\phi,\pi)};\xi)$ is known in random matrix theory
as the generating function of the gap probabilities.  Here CUE
stands for {\it Circular Unitary Ensemble}, which, as already
noted, denotes the probability space given by $\UN$ equipped with
the Haar measure. Differentiating one obtains
\begin{equation}
\EgU(n;I_{[\pi -\phi,\pi)}) = (-1)^n\left. \frac{\rmd^n
\EgU(I_{[\pi-\phi,\pi)};\xi)}{\rmd \xi^n} \right |_{\, \xi=1},
\end{equation}
where $\EgU(n;I_{[\pi-\phi,\pi)})$ is the probability that the
interval $I_{[\pi-\phi,\pi)}$ contains $n$ eigenvalues.  Making
the substitution
\begin{equation}
\rme^{\rmi \theta} = \frac{1 + \rmi x}{1 - \rmi x}, \quad x = \tan
\frac\theta2
\end{equation}
equation~\eqref{fdef} becomes
\begin{equation}
\label{cvar}
\begin{split}
\EgC(s;\xi)&=\frac{2^{N^2}}{(2\pi)^N
N!}\int_{-\infty}^{\infty} \cdots
\int_{-\infty}^{\infty}\prod_{j=1}^N \frac{\left[1 - \xi
\chi_{I_{[s, \infty)}}(x_j)\right]}{\left(1 + x_j^2\right)^N}\,\\
& \quad \times
\prod_{1\leq j < k \leq N}\left|x_j - x_k\right|^2 \rmd x_1\cdots
\rmd x_N,
\end{split}
\end{equation}
where $s = \cot(\phi/2)$.  The integral~\eqref{cvar} is the gap
probability generating function of the Cauchy ensemble (CyUE) for
the interval $[s,\infty)$.

If we let the exponent in the denominator in the
integrand~\eqref{cvar} vary, we can define
\begin{equation}
\begin{split}
 \EgC(s;\eta,\xi)&=\frac{2^{N^2}}{(2\pi)^N
N!}\int_{-\infty}^{\infty} \cdots
\int_{-\infty}^{\infty}\prod_{j=1}^N \frac{\left[1 - \xi
\chi_{I_{[s, \infty)}}(x_j)\right]}{\left(1 + x_j^2\right)^\eta}\,
\\
& \quad \times \prod_{1\leq j < k \leq N}\left|x_j - x_k\right|^2 \rmd x_1\cdots
\rmd x_N.
\end{split}
\end{equation}
The connection between the gap probability generating function of
$\UN$ and the theory of Painlev\'e equations is mediated by  the
function
\begin{equation}
\label{Painvsol} \sigma(s)=(1 + s^2)\frac{\rmd}{\rmd s} \ln
\EgC(s;a +N,\xi),
\end{equation}
which is a solution of the equation
\begin{gather}
(1 + s^2)^2(\sigma'')^2 + 4(1 + s^2)(\sigma')^3
-8s\sigma(\sigma')^2 + 4\sigma^2(\sigma' -a^2)\nonumber \\
\label{Painvvi}
 + 8a^2 s \sigma
\sigma' + 4[N(N+2a) -a^2s^2](\sigma')^2=0.
\end{gather}
The above equation is known in the literature as the $\sigma$-form
of the Painlev\'e VI equation.

Solutions of equation~\eqref{Painvvi}, and therefore the
determinant~\eqref{fdef}, obey a recurrence relation which allows
one to determine an exact formula for them for each value of $N$.
Therefore, at least in principle, it is possible to compute the
entropy of the entanglement~\eqref{JKidea2} exactly for each $N$.
In terms of of the gap probability generating function, these
recurrence relations assume the following form~\cite{FW03}:
\begin{equation}
\label{rec1}
\frac{E^{\mathrm{CUE}}_{N-1}
E^{\mathrm{CUE}}_{N+1}}{(E^{\mathrm{CUE}}_{N})^2} = 1 - x_N^2
\end{equation}
with initial conditions
\begin{equation}
\label{incon1} E_0^{\mathrm{CUE}} = 1 \quad E_1^{\mathrm{CUE}} = 1
- \frac{\xi}{2\pi} \phi.
\end{equation}
In turn $x_N$ obeys the recurrence relation
\begin{equation}
\label{rec2}
\begin{split}
2x_N x_{N-1} - 2\cos \frac{\phi}{2}& =
\frac{1-x_N^2}{x_N}\left[(N+1)x_{N+1} - (N-1)x_{N-1}\right] \\
& \quad - \frac{1-x_{N-1}^2}{x_{N-1}^2}\left[Nx_N -
(N-2)x_{N-2}\right]
\end{split}
\end{equation}
with initial conditions
\begin{equation}
\label{incon2}
x_{-1} =0,\quad x_0 = 1\quad {\rm and} \quad x_1
=-\frac{\xi}{\pi}\frac{\sin\frac{\phi}{2}}{1-\frac{\xi}{2\pi}\phi}.
\end{equation}

These recurrence relations can be used to compute higher order
terms in the asymptotics of $E^{\mathrm{CUE}}_N$ as $N \rightarrow
\infty$, and therefore of the entropy of the entanglement.
Substituting the Fisher-Hartwig formula~\eqref{FHconj}
into~\eqref{rec1} gives
\begin{equation}
x_N \sim \frac{\sqrt{2}\abs{\beta(\lambda)}}{N}, \quad N
\longrightarrow \infty,
\end{equation}
where $\beta(\lambda)$ was defined in~\eqref{chc}.  This suggests
that $x_N$ has an asymptotic expansion of the form
\begin{equation}
x_N \sim \frac{\sqrt{2}\abs{\beta(\lambda)}}{N} +
\frac{c_1(\lambda)}{N^2} + \frac{c_2(\lambda)}{N^3} + \cdots \quad
N \rightarrow \infty.
\end{equation}
Inserting this expansion into~\eqref{rec2} it is possible to
compute recursively the coefficients $c_j(\lambda)$; for the
second order term we obtain
\begin{equation}
c_2(\lambda) = 2^{1/3}\abs{\beta(\lambda)}^3.
\end{equation}
This coefficient determines the contribution to the
integral~\eqref{JKidea2} given by equation~\eqref{O1term}.  Higher
order terms can be computed in a similar way.

When the average is over the other compact groups, the
determinant~\eqref{chpoln} can still be interpreted as gap
probability generating function for the respective group and its
logarithm is still a solution of a differential equation related
to the Painlev\'e VI equation.  However, although recurrence
formulae analogous to~\eqref{rec1} exist, at each step the values
of $\lamp$ that label the integral~\eqref{avsh} change and in
general do not even identify one of the classical compact groups
(see, e.g.,~\cite{FW02}).

\section{Conclusions}
\label{conclusions}

We have investigated the entanglement of formation of the ground
state of the general class of quantum spin chains related to
quadratic Hamiltonians of fermionic oscillators partitioned into
two contiguous subchains. The number of oscillators in the first
subsystem is $N$ and we let the total size of the system grow to
infinity.  We have discovered that for certain Hamiltonians the
measure of entanglement, which in these circumstances is given by
the von Neumann entropy of the first subchain, can be expressed in
term of an average over one of the classical compact groups $\UN$,
$\SpN$ and $\mathrm{O}^\pm(N)$. Indeed, we show that there exists
a one-to-one correspondence between the symmetries of the
fermionic chain and the functional form of the Haar measures of
the classical compact groups. Using generalizations of the
Fisher-Hartwig conjecture it is possible to compute asymptotic
formulae for such averages in the limit $N\rightarrow \infty$. The
entanglement is either zero, away from critical point, or grows
logarithmically with $N$ in the proximity of quantum phase
transitions.  Generalizations of the Fisher-Hartwig formula allow
one to determine the constant in front of the logarithm
explicitly, and the next-to-leading-order term in the asymptotics.
Furthermore, these averages turn out to be related to solutions of
Painlev\'e equations.

The fact that one can compute the leading order terms in
 the asymptotics of the entanglement of formation of the ground state
of such a significant class of systems suggests that the
connection between lattice models and random matrices may be
deeper than being simply a calculational device. For example, the
diagonal spin-spin correlations of the two-dimensional classical
Ising model are Toeplitz determinants with symbols analogous to
the one associated to the XY model; it is likely that a similar
association between classical compact groups and symmetries of the
Hamiltonian exists also for classical lattice models.  If it does,
random-matrix techniques could then be used to deduce
thermodynamic quantities like critical exponents. After all,
two-dimensional classical spin chains are mathematically
equivalent to one-dimensional quantum spin chains. Furthermore,
these type of random matrix averages already appear in the
calculation of the ground state density matrices for an
impenetrable Bose gas in an interval of finite length~\cite{FF04}.

It is likely that these applications of group averages will prompt
further studies of the analytical properties of the determinants
and spectra of combination of Toeplitz and Hankel matrices, the
investigation of which has started only recently.

\section*{Acknowledgments}
We gratefully acknowledge stimulating discussions with Estelle
Basor, Peter Forrester and Noah Linden. We are also grateful for
the kind hospitality of the Isaac Newton Institute for the
Mathematical Sciences, Cambridge, while this research was
completed. Francesco Mezzadri was supported by a Royal Society
Dorothy Hodgkin Research Fellowship.

\appendix

\section*{Appendix A. The density matrix of a subchain}
\renewcommand{\theequation}{A.\arabic{equation}}

Let $\{\ket{\psi_j}\}$ be a basis of the Hilbert space
$\mathcal{H}$ of a system composed of two parts, P and Q, so that
$\mathcal{H}=\mathcal{H}_{\mathrm{P}} \otimes
\mathcal{H}_{\mathrm{Q}}$.  The density matrix of a statistical
ensemble expressed in the basis $\{\ket{\psi_j}\}$ is a positive
Hermitian matrix given by
\begin{equation}
\rho_{\mathrm{PQ}} = \sum_{jk}
c_{jk}\ket{\psi_j}\negthinspace\bra{\psi_k},
\end{equation}
with the condition $\trace_{\mathrm{PQ}} \rho_{\mathrm{PQ}} =1$.
Let us introduce the operators $S(j,k)$ and $\overline{S}(j,k)$
defined by the relations
\begin{subequations}
\begin{gather}
 S(j,k)  = \ket{\psi_j}\negthinspace\bra{\psi_k} \\
 \overline{S}(j,k)S(k,l) = \delta_{jl}\ket{\psi_j}\negthinspace\bra{\psi_j}
 \quad \text{and} \quad S(j,k)\overline{S}(k,l) =
 \delta_{jl}\ket{\psi_j}\negthinspace\bra{\psi_j}.
 \end{gather}
 \end{subequations}
(In this formula repeated indices are not summed over.)  Clearly,
 we have
\begin{equation}
c_{jk} = \trace_{\mathrm{PQ}}\left[\rho_{\mathrm{PQ}}
\,\overline{S}(k,j)\right].
\end{equation}

Let us now suppose that the Hamiltonian of our physical system
is~\eqref{impH} and that the subsystem P is composed of the first
$N$ oscillators. Then a set of operators $S(j,k)$ for the subchain
P can be generated by products of the type $\prod_{j=1}^NG_j$,
where $G_j$ can be any of the operators
$\{c_j,c^\dagger_j,c^\dagger_jc_j,c_jc_j^\dagger \}$ and the
$c_j$s are Fermi operators that span $\mathcal{H}_{\mathrm{P}}$;
it is straightforward to check that $\overline{S}(k,j) =
\left(\prod G_{j=1}^N\right)^\dagger$. We then have
\begin{equation}
\begin{split}
\rho_{\mathrm{P}} & = \sum_{\text{All the $S(l,k)$}}\trace_{{\rm
P}}\left[\rho_{\mathrm{P}}\left(\prod_{j=1}^N
G_j\right)^\dagger\right]\prod_{j=1}^N G_j \\
&= \sum_{\text{All the $S(l,k)$}}\trace_{{\rm
P}}\left[\trace_{\mathrm{Q}}\left(\rho_{\mathrm{PQ}}\right)\left(\prod_{j=1}^N
G_j\right)^\dagger\right]\prod_{j=1}^N G_j \\
& = \sum_{\text{All the $S(l,k)$}}\trace_{{\rm
PQ}}\left[\rho_{\mathrm{PQ}}\left(\prod_{j=1}^N
G_j\right)^\dagger\right]\prod_{j=1}^N G_j.
\end{split}
\end{equation}
Since $\rho_{\mathrm{PQ}}=\gsk\negthinspace \gsb$, this expression
simply reduces to
\begin{equation}
\rho_{\mathrm{P}}=\sum_{\text{All the $S(l,k)$}}\Biggl \langle
\boldsymbol{\Psi}_{\mathrm{g}} \Biggr | \left(\prod_{j=1}^N
G_j\right)^\dagger \Biggr |\boldsymbol{\Psi}_{\mathrm{g}} \Biggr
\rangle \prod_{j=1}^N G_j.
\end{equation}
The correlation functions in the above sum can be computed using
Wick's theorem~\eqref{Wick-Th}. Finally, if the correlations of
the $c_j$s are given by~\eqref{exval}, we immediately obtain
formula~\eqref{rop}.

\section*{Appendix B. The correlation matrix $C_M$}
\renewcommand{\theequation}{B.\arabic{equation}}
\label{corrmatap} The purpose of this appendix is to provide an
explicit derivation of the expectation values
\begin{equation}
\label{objective}
\gsb m_jm_k \gsk
\end{equation}
when the dynamics is determined by the Hamiltonian~\eqref{impH}.

First, we need to diagonalize $H_{\alpha}$, which is achieved by
finding  a linear transformation of the operators $b_j$ of the
form
\begin{equation}
\label{lintras}
\eta_k = \sum_{j=0}^{M-1}\left(g_{kj} b_j + h_{kj}
b_j^\dagger\right),
\end{equation}
 such that the Hamiltonian~\eqref{impH} becomes
\begin{equation}
\label{diagH}
H_{\alpha} = \sum_{k=0}^{M-1}\abs{\Lambda_k}\,
\eta_k^\dagger\eta_k + C,
\end{equation}
where the coefficients $g_{kj}$ and $h_{kj}$ are real, the
$\eta_k$s are Fermi operators and $C$ is a constant. We use the
notation $\abs{\Lambda_k}$ because it is convenient for the
computations carried out in section~\ref{unitary} to allow the
coefficients of the number operators $\eta^\dagger_k\eta_k$ to be
the absolute values of the complex numbers $\Lambda_k$. The
quadratic form~\eqref{impH} can be transformed into~\eqref{diagH}
by~\eqref{lintras} if the system of equations
\begin{equation}
\label{eigeq1}
 [\eta_k, H_{\alpha}] - \abs{\Lambda_k} \eta_k =0, \quad
k=0,\ldots, M-1
\end{equation}
has a solution.  Substituting~\eqref{impH} and~\eqref{lintras}
into~\eqref{eigeq1} we obtain the eigenvalue equations
\begin{subequations}
\label{eigeq2}
\begin{align}
\abs{\Lambda_k} g_{kj} & = \sum_{l=0}^{M-1}\left(g_{kl}\bA_{lj} -
h_{kl}\bB_{lj}\right), \\
\abs{\Lambda_k} h_{kj} & = \sum_{l=0}^{M-1}\left(g_{kl}\bB_{lj} -
h_{kl}\bA_{lj}\right),
\end{align}
\end{subequations}
where $\bA = \alpha A - 2 I$ and $\bB =\alpha \gamma B$. These
equations can be simplified by setting
\begin{subequations}
\label{subst}
\begin{align}
\phi_{kj}& = g_{kj} + h_{kj}  \\
\psi_{kj} & = g_{kj}-h_{kj},
\end{align}
\end{subequations}
in terms of which the equations~\eqref{eigeq2} become
\begin{subequations}
\label{fstep}
\begin{align}
\label{phi1}
(\bA+\bB)\vphi_k &= \abs{\Lambda_k} \vpsi_k \\
\label{psi1} (\bA - \bB)\vpsi_k &= \abs{\Lambda_k}\vphi_k.
\end{align}
\end{subequations}
Combining these two expressions, we obtain
\begin{subequations}
\label{step}
\begin{align}
\label{phi2}
(\bA-\bB)(\bA+\bB)\vphi_k & = \abs{\Lambda_k}^2\vphi_k \\
\label{psi2} (\bA+\bB)(\bA-\bB)\vpsi_k  & =
\abs{\Lambda_k}^2\vpsi_k.
\end{align}
\end{subequations}
When $\Lambda_k \neq 0$, $\vphi_k$ and $\abs{\Lambda_k}$ can be
determined by solving the eigenvalue equation~\eqref{phi2}, then
$\vpsi_k$ can be computed using~\eqref{phi1}. Alternatively, one
can solve equation~\eqref{psi2} and then obtain $\vphi_k$
from~\eqref{psi1}. When $\Lambda_k=0$, $\vphi_k$ and $\vpsi_k$
differ at most by a sign and can be deduced directly either
from~\eqref{fstep} or from~\eqref{step}.

Since $\bA$ and $\bB$ are real, the matrices $(\bA-\bB)(\bA +\bB)$
and $(\bA+\bB)(\bA-\bB)$ are symmetric and positive, which
guarantees that all of their eigenvalues are positive.
Furthermore, the $\vphi_k$s and $\vpsi_k$s can be chosen to be
real and orthonormal. As a consequence the coefficients $g_{kj}$
and $h_{kj}$ obey the constraints
\begin{subequations}
\begin{gather}
\sum_{k=0}^{M-1}\left(g_{kj}g_{kl} + h_{kj}h_{kl}\right) =
\delta_{jl}, \\
\sum_{k=0}^{M-1}\left(g_{kj}h_{kl} + h_{kj}g_{kl}\right) =  0,
\end{gather}
\end{subequations}
which are necessary and sufficient conditions for the $\eta_k$s to
be Fermi operators.

The constant in equation~\eqref{diagH} can be computed by taking
the trace of $H_{\alpha}$ using the two expressions~\eqref{impH}
and~\eqref{diagH}:
\begin{equation}
\trace H_{\alpha}= 2^{M-1}\sum_{k=0}^{M-1} \left(\alpha A_{kk} -
2\right) = 2^{M-1}\sum_{k=0}^{M-1} \abs{\Lambda_k} + 2^M C.
\end{equation}
Therefore, we have
\begin{equation}
C = \frac{1}{2}\sum_{k=0}^{M-1}\left(\alpha A_{kk} - 2 -
\abs{\Lambda_k}\right).
\end{equation}

We are now in a position to compute the contraction
pair~\eqref{objective}.  Substituting~\eqref{subst}
into~\eqref{lintras} we have
\begin{equation}
\label{neta} \eta_k = \frac{1}{2}\sum_{j=0}^{M-1}
\left(\phi_{kj}m_{2j + 1} - \rmi\psi_{kj}m_{2j}\right).
\end{equation}
Since the $\vphi_k$s and $\vpsi_k$s are two sets of real and
orthogonal vectors,~\eqref{neta} can be inverted to give
\begin{subequations}
\label{mexp}
\begin{align}
m_{2j} & = \rmi \sum_{k=0}^{M-1} \psi_{kj}\left(\eta_k -
\eta_k^\dagger\right) \\
m_{2j + 1} &=\sum_{k=0}^{M-1} \phi_{kj}\left(\eta_k +
\eta_k^\dagger\right).
\end{align}
\end{subequations}
Since the vacuum state of the operators $\eta_k$ coincides with
$\gsk$, the expectation values~\eqref{objective} are easily
computed from the expressions~\eqref{mexp}. We have
\begin{subequations}
\label{coelc}
\begin{equation}
\begin{split}
\gsb m_{2j}m_{2k} \gsk & =
\sum_{l=0}^{M-1}\psi_{lj}\psi_{lk}=\delta_{jk}, \\
 \gsb m_{2j+1}m_{2k+1} \gsk & =
\sum_{l=0}^{M-1}\phi_{lj}\phi_{lk}=\delta_{jk}
\end{split}
\end{equation}
and
\begin{equation}
\begin{split}
\gsb m_{2j}m_{2k + 1} \gsk &= \rmi
\sum_{l=0}^{M-1}\psi_{lj}\phi_{lk}, \\
\gsb
m_{2j+1}m_{2k} \gsk & = -\rmi \sum_{l=0}^{M-1}\psi_{lk}\phi_{lj}.
\end{split}
\end{equation}
\end{subequations}
Finally, by introducing the real $M \times M$ matrix
\begin{equation}
\label{Tmat}
\left(T_M\right)_{jk} = \sum_{l=0}^{M-1}\psi_{l
j}\phi_{lk}, \quad j,k=0,\ldots, M-1
\end{equation}
and combining the expressions~\eqref{coelc} we obtain
\begin{equation}
\gsb m_jm_k \gsk = \delta_{jk} + \rmi (C_M)_{jk},
\end{equation}
where the matrix $C_M$ has the block structure
\begin{subequations}
\label{corrmatg}
\begin{equation}
C_M = \begin{pmatrix} C_{11} & C_{12} & \cdots & C_{1M} \\
                      C_{21} & C_{22} & \cdots & C_{2M} \\
                      \hdotsfor[2]{4} \\
                      C_{M1} & C_{M2} & \cdots & C_{MM}
\end{pmatrix}
\end{equation}
with
\begin{equation}
\label{blockC2}
C_{jk} = \begin{pmatrix} 0 & (T_M)_{jk} \\
                         -(T_M)_{kj} & 0
         \end{pmatrix}.
\end{equation}
\end{subequations}
We call $C_M$ the correlation matrix.  It is worth noting that
because of the definition~\eqref{Tmat}, the matrix $T_M$ contains
all of the physical information relating to the ground state of
$H_{\alpha}$.

\section*{Appendix C. Averages over the classical compact groups}
\label{cgroupav}

\renewcommand{\theequation}{C.\arabic{equation}}
Let $f(\theta)$ be a $2\pi$-periodic even function and let $\GM$
be one of the classical compact groups $\ONep$, $\ONen$, $\SptN$,
${\rm O}^+(2N+1)$ and ${\rm O}^{-}(2N + 1)$, where the superscript
$\pm$ denotes the connected component of the orthogonal group with
positive and negative determinant respectively. The integer
$\tilde{N}$ denotes the total number of eigenvalues, while $N$
denotes the number of independent ones.  For any $U \in \GM$, let
$F(U)$ be the class function defined by
\begin{equation}
F(U)=\prod_{j=1}^{\tilde N} f(\theta_j).
\end{equation}
In this appendix we want to compute averages of the type
\begin{equation}
\label{cgav} \Bigl \langle F(U)\Bigr \rangle_{\GM} = \left \langle
\prod_{j=1}^{\tilde N} f(\theta_j) \right\rangle_{\GM}.
\end{equation}

The Haar measure of $\GM$ expressed in terms of the independent
eigenvalues is given by (see, e.g.,~\cite{Wey46}, pp. 218 and 224)
\begin{equation}
\label{HmeasW}
\begin{split}
P^N_{(\sigma_1,\sigma_2)}(\theta_1,\ldots,\theta_N) &
=\frac{1}{Z_N^{(\sigma_1,\sigma_2)}}\prod_{l=1}^N(1 +
\cos\theta_l)^{\sigma_1 + 1/2}(1 - \cos \theta_l)^{\sigma_2 +
1/2}\\
& \quad \times \prod_{1 \le  j < k \le N} (\cos\theta_j -
\cos\theta_k)^2,
\end{split}
\end{equation}
 where $Z_N^{(\sigma_1,\sigma_2)}$ is a normalization constant whose specific
 value will not be relevant to what follows. The parameters
$(\sigma_1,\sigma_2)=(-1/2,-1/2)$, $(1/2,1/2)$, $(-1/2,1/2)$,
$(1/2,-1/2)$ refer to $\ONep$, $\SptN$, ${\rm O}^+(2N+1)$ and
${\rm O}^{-}(2N + 1)$ respectively; we will consider $\ONen$
separately.

Averages of the type~\eqref{cgav} can be evaluated by making the
substitutions
\begin{equation}
\label{sub} x_j = \cos \theta_j,
\end{equation}
which reduces~\eqref{cgav} to the computation of the integral
\begin{equation}
\label{intort}
\begin{split}
I_N(\sigma_1,\sigma_2)& = \frac{1}{\nZ}\int_{-1}^1 \cdots
\int_{-1}^1\left(\prod_{j=1}^N g(\cos^{-1}x)\right)\prod_{j=1}^N
\left(1+x_j\right)^{\sigma_1}\left(1-x_j\right)^{\sigma_2} \\
& \quad \times\prod_{1\le j <k \le N}\left(x_j - x_k\right)^2\rmd
x_1\cdots \rmd x_N,
\end{split}
\end{equation}
where $g(\theta)=f(\theta)f(-\theta)$.  For $\sigma_1,\sigma_2 >
-1$ the integral~\eqref{intort} can be evaluated using orthogonal
polynomial techniques; the details of the computation can be
found, for example, in sections~2.1 and~2.2 of~\cite{Seg59}.  Let
us introduce the weight
\begin{equation}
\label{weight} w_{(\sigma_1,\sigma_2)}(x)=(1 +
x)^{\sigma_1}(1-x)^{\sigma_2}, \quad -1< x < 1
\end{equation}
together with the set of polynomials
$\{p_j^{(\sigma_1,\sigma_2)}(x)\}$ orthogonal with respect to
$w_{(\sigma_1,\sigma_2)}$, i.e.
\begin{equation}
\int_{-1}^1 w_{(\sigma_1,\sigma_2)}(x)p^{(\sigma_1,\sigma_2)}_j(x)
p^{(\sigma_1,\sigma_2)}_k(x)\rmd x = \delta_{jk}.
\end{equation}
We have
\begin{equation}
I_N(\sigma_1,\sigma_2) =
\det(\alpha^{(\sigma_1,\sigma_2)}_{jk})_{j,k=0,\ldots,N-1},
\end{equation}
where
\begin{equation}
\label{ortmel} \alpha^{(\sigma_1,\sigma_2)}_{jk}=\int_{-1}^1
g(\cos^{-1}x)w_{(\sigma_1,\sigma_2)}(x)p^{(\sigma_1,\sigma_2)}_j(x)
p^{(\sigma_1,\sigma_2)}_k(x)\rmd x.
\end{equation}
The orthogonal polynomials $\{p_j^{(\sigma_1,\sigma_2)}(x)\}$ are
called {\it Jacobi polynomials}.

\begin{enumerate}
\item  $\ONep$. The average to compute is
\begin{equation}
\begin{split}
\left \langle \prod_{j=1}^{\tilde N} f(\theta_j)
\right\rangle_{\ONep}& =\int_{-\pi}^\pi \cdots \int_{-\pi}^{\pi}
\left(\prod_{j=1}^N
f(\theta_j)f(-\theta_j)\right)\\
& \quad \times P^N_{(-1/2,-1/2)}(\theta_1,\ldots,\theta_N)
\rmd \theta_1 \cdots \rmd \theta_N.
\end{split}
\end{equation}
The substitution~\eqref{sub} gives
\begin{equation}
\left \langle \prod_{j=1}^{\tilde N} f(\theta_j)
\right\rangle_{\ONep}= I_N(-1/2,-1/2) =
\det\left(\alpha_{jk}\right)_{j,k=0,\ldots,N-1}.
\end{equation}
(For simplicity, from now on we shall drop the superscript
$(\sigma_1,\sigma_2)$ when denoting the orthogonal polynomials,
their weight and the matrix elements~\eqref{ortmel}.)
 When the weight is $w(x)=(1-x^2)^{-1/2}$ the Jacobi polynomials are
 also known as Chebyshev polynomials of the first kind; these are
\begin{equation}
\label{Chpolyft}
p_0(x) = \frac{1}{\sqrt{\pi}} \quad {\rm and }
\quad p_j(x)=\sqrt{\frac{2}{\pi}}\cos\left(j\cos^{-1}x\right),
\quad j
>0.
\end{equation}
Substituting $x = \cos \theta$ in equation~\eqref{ortmel} we have
\begin{subequations}
\label{inttro2n}
\begin{align}
\alpha_{00} &= \frac{1}{2\pi} \int_0^{2\pi} g(\theta) \rmd \theta
=g_0,
\\
\alpha_{0j}& =\alpha_{j0} = \frac{\sqrt{2}}{\pi}\int_0^\pi
g(\theta)\cos( j\, \theta ) \rmd \theta = \sqrt{2}g_j, \quad j> 0, \\
\alpha_{jk} & = \frac{2}{\pi} \int_0^\pi g(\theta)
\cos(j\,\theta)\cos(k\, \theta)\rmd \theta = g_{j-k} + g_{j + k},
\quad j,k > 0
\end{align}
\end{subequations}
where $g_j = \frac{1}{2\pi}\int_0^{2\pi} g(\theta)\rme^{-\rmi j \,
\theta}\rmd \theta$. \item $\SptN$. The average to compute is
\begin{equation}
\begin{split}
\left \langle \prod_{j=1}^{\tilde N} f(\theta_j)
\right\rangle_{\SptN}& =\int_{-\pi}^\pi \cdots \int_{-\pi}^{\pi}
\left(\prod_{j=1}^N
f(\theta_j)f(-\theta_j)\right) \\
& \quad \times P^{N}_{(1/2,1/2)}(\theta_1,\ldots,\theta_N)
\rmd \theta_1 \cdots \rmd \theta_N.
\end{split}
\end{equation}
The substitution~\eqref{sub} gives
\begin{equation}
\left \langle \prod_{j=1}^{\tilde N} f(\theta_j)
\right\rangle_{\SpN}= I_N(1/2,1/2)=
\det\left(\alpha_{jk}\right)_{j,k=0,\ldots,N-1}.
\end{equation}
When the weight is $w(x)=(1-x^2)^{1/2}$ the Jacobi polynomials
reduce to the Chebyshev polynomials of the second kind:
\begin{equation}
p_j(x) = \sqrt{\frac{2}{\pi}} \frac{\sin\left[(j + 1)\cos^{-1}
x\right]}{\sin\left(\cos^{-1} x \right)},\quad j\ge0.
\end{equation}
The matrix elements~\eqref{ortmel} becomes
\begin{equation}
\label{symal}
\begin{split}
\alpha_{jk} & = \frac{2}{\pi}\int_0^\pi
g(\theta)\sin\left[(j+1)\theta\right]\sin\left[(k+1)\theta\right]\rmd
\theta \\
& = g_{j-k} - g_{j + k + 2}, \quad j,k=0,\ldots,N -1.
\end{split}
\end{equation}

\item $\ONo{+}$.  Since the determinant is positive and the
dimension is odd, the eigenvalues come in complex conjugate pairs
and the extra eigenvalue is $1$. This property must be taken into
account when computing the average~\eqref{cgav}.  We have the
formula
\begin{equation}
\begin{split}
\left \langle \prod_{j=1}^{\tilde N} f(\theta_j)
\right\rangle_{\ONo{+}}&=f(0)\int_{-\pi}^\pi \cdots
\int_{-\pi}^{\pi} \left(\prod_{j=1}^N
f(\theta_j)f(-\theta_j)\right)\\
& \quad \times P^{N}_{(-1/2,1/2)}(\theta_1,\ldots,\theta_N) \rmd
\theta_1 \cdots \rmd \theta_N,
\end{split}
\end{equation}
which can be rearranged to give
\begin{equation}
\left \langle \prod_{j=1}^{\tilde N} f(\theta_j)
\right\rangle_{\ONo{+}}=
f(0)I_N(-1/2,1/2)=f(0)\det\left(\alpha_{jk}\right)_{j,k=0,\ldots,N-1}.
\end{equation}
The Jacobi polynomials with weight $w(x)=\sqrt{(1-x)/(1+x)}$ are
\begin{equation}
p_j(x) = \frac{1}{\sqrt{\pi}}\frac{\sin\left[\left(\frac{2j +
1}{2}\right)\cos^{-1}
(x)\right]}{\sin\left[\frac12\cos^{-1}(x)\right]}, \quad j\ge 0.
\end{equation}
The matrix elements~\eqref{ortmel} become
\begin{equation}
\begin{split}
\alpha_{jk}&
=\frac{2}{\pi}\int_0^{\pi}g(\theta)\sin\left[\left(\frac{2j
+1}{2}\right)\theta\right]\sin\left[\left(\frac{2k +
1}{2}\right)\theta\right]\rmd \theta \\
&=g_{j-k} - g_{j + k + 1},\quad j,k=0,\ldots,N-1.
\end{split}
\end{equation}
\item $\ONo{-}$.  The treatment in this case is similar to that
one for $\ONo{+}$, except that now the extra eigenvalue is $-1$.
Therefore, we have
\begin{equation}
\left \langle \prod_{j=1}^{\tilde N} f(\theta_j)
\right\rangle_{\ONo{-}}=
f(\pi)I_N(1/2,-1/2)=f(\pi)\det\left(\alpha_{jk}\right)_{j,k=0,\ldots,N-1}.
\end{equation}
The weight is $w(x)=\sqrt{\left(1 + x\right)/\left(1-x\right)}$
and the corresponding Jacobi polynomials are
\begin{equation}
p_j(x) = \frac{1}{\sqrt{\pi}}\frac{\cos\left[\left(\frac{2j +
1}{2}\right)\cos^{-1}
(x)\right]}{\cos\left[\frac12\cos^{-1}(x)\right]},\quad j\ge 0.
\end{equation}
The matrix elements~\eqref{ortmel} are given by the integral
transform
\begin{equation}
\begin{split}
\alpha_{jk}&
=\frac{2}{\pi}\int_0^{\pi}g(\theta)\cos\left[\left(\frac{2j
+1}{2}\right)\theta\right]\cos\left[\left(\frac{2k +
1}{2}\right)\theta\right]\rmd \theta \\
& =g_{j-k} + g_{j + k + 1},\quad j,k=0,\ldots,N-1.
\end{split}
\end{equation}

\begin{table}
\label{tabongroups} \centering
\begin{tabular}{||c|c|l||}
\hline \hline $\GM$ & $\left \langle \prod_{j=1}^{\tilde N}
f(\theta_j) \right \rangle_{\GM}$ & \hspace{2.65 cm} $\alpha^{(\sigma_1,\sigma_2)}_{jk}$ \\
\hline $\UN$ & $\det\left(\alpha_{jk}\right)_{j,k=0,\ldots,N-1}$
& $f_{j-k}, \quad j,k \ge 0$ \\
 $ \ONep$ & $\det\left(\alpha_{jk}\right)_{j,k=0,\ldots,N-1}$ &
\begin{tabular}{c l}
  $g_0$ & if $j=k=0$ \\
  $\sqrt{2}g_l$ & if either $j=0$,  $k=l$ \\
                & or $j=l$, $k=0$\\
  $g_{j-k} + g_{j+k}$ & if $j,k >0$
\end{tabular} \\
$\SptN$ &  $\det\left(\alpha_{jk}\right)_{j,k=0,\ldots,N-1}$ &
$g_{j-k} -  g_{j+k +2}, \quad j,k \ge 0$ \\
 $\ONo{+}$ &
$f(0)\det\left(\alpha_{jk}\right)_{j,k=0,\ldots,N-1}$ & $g_{j-k} -
g_{j+k+1}, \quad j,k \ge 0$\\  $\ONo{-}$ &
$f(\pi)\det\left(\alpha_{jk}\right)_{j,k=0,\ldots,N-1}$ & $g_{j-k}
+ g_{j+k+1}, \quad j,k \ge 0$\\ $\ONen$ &
$f(0)f(\pi)\det\left(\alpha_{jk}\right)_{j,k=0,\ldots,N-1}$ &
$g_{j-k} -  g_{j+k +2}, \quad j,k \ge 0$ \\ \hline \hline
\end{tabular}
\caption{Summary of the averages over the different compact groups
of $F(U) = \prod_{j=1}^{\tilde N} f(\theta_j)$. The function $f$
is even and $g(\theta) = f(\theta)f(-\theta)$; $f_l$ and $g_l$ are
the corresponding Fourier coefficients; $\tilde N$ is the
dimension of the matrices and $N$ the number of independent
eigenvalues.}
\end{table}

\item $\ONen$.  Since the determinant is negative, one eigenvalue
must be $-1$, and since they come in complex conjugate pairs $1$
is also an eigenvalue.  The number of independent eigenvalues is
therefore only $N$ and not $N+1$.  The Haar measure is given by
formula~\eqref{HmeasW} with $(\sigma_1,\sigma_2)=(1/2,1/2)$, the
same as $\SptN$. Therefore, we have
\begin{equation}
\begin{split}
\left \langle \prod_{j=1}^{\tilde N} f(\theta_j)
\right\rangle_{\ONen}& =f(0)f(\pi)\int_{-\pi}^\pi \cdots
\int_{-\pi}^{\pi} \left(\prod_{j=1}^N
f(\theta_j)f(-\theta_j)\right)\\
& \quad \times P^{N}_{(1/2,1/2)}(\theta_1,\ldots,\theta_N) \rmd
\theta_1 \cdots \rmd \theta_N,
\end{split}
\end{equation}
which becomes
\begin{equation}
\begin{split}
\left \langle \prod_{j=1}^{\tilde N} f(\theta_j)
\right\rangle_{\ONen} &
=f(0)f(\pi)I_N(1/2,1/2)\\
& =f(0)f(\pi)\det\left(\alpha_{jk}\right)_{j,k=0,\ldots,N-1},
\end{split}
\end{equation}
where the $\alpha_{jk}$ are given by equation~\eqref{symal}.
\end{enumerate}

There exist useful relations among averages of functions over the
classical compact groups. Those that are relevant to this paper
are the following:
\begin{subequations}
\begin{align}
\left \langle \prod_{j=1}^{2N +1} g(\theta_j)\right \rangle_{{\rm
U}(2N + 1)}&=\left \langle \prod_{j=1}^{2N + 2} f(\theta_j)\right
\rangle_{{\rm O}^{+}(2N +2)}\left \langle \prod_{j=1}^{2N}
f(\theta_j)\right \rangle_{\SpN},\\
 \label{ONeav} \left \langle
\prod_{j=1}^{\tilde N} f(\theta_j)\right \rangle_{\ONo{+}} &=
\left \langle \prod_{j=1}^{\tilde N} f(\pi - \theta_j)\right
\rangle_{\ONo{-}},\\  
\label{ONeavuf} \left \langle \prod_{j=1}^{\tilde N}
f(\theta_j)\right \rangle_{\ONo{-}} & = \left \langle
\prod_{j=1}^{\tilde N} f(\pi - \theta_j)\right \rangle_{\ONo{+}}, \\
\label{leq}
 \left \langle
\prod_{j=1}^{2N}f(\theta_j) \right \rangle_{\SptN} & =
\frac{1}{f(0)f(\pi)}  \left \langle \prod_{j=1}^{2N+2}f(\theta_j)
\right \rangle_{\ONen}.
\end{align}
\end{subequations}

\end{document}